\pdfoutput=1
\documentclass{article}
\usepackage[utf8]{inputenc}%
\usepackage[tracking=true, letterspace=100, expansion=false]{microtype}
\usepackage[T1]{fontenc}
\usepackage[title]{appendix}%

\usepackage{tabularx}%
\usepackage{siunitx}
\usepackage{booktabs}%
\usepackage[normal, online, flushleft]{threeparttable}%

\usepackage{pdflscape}%

\usepackage{amsmath}%
\usepackage{amsfonts}%
\usepackage{amsthm}%

\newtheorem{theorem}{Theorem}
\theoremstyle{definition}

\theoremstyle{remark}

\usepackage[margin=1.2in]{geometry}
\usepackage[onehalfspacing]{setspace}
\usepackage{mathtools}
\DeclarePairedDelimiter\indicatorfence{\{}{\}}
\DeclarePairedDelimiter\abs{\lvert}{\rvert}

\newcommand\1{\operatorname{\mathbb{I}}\indicatorfence}

\newcommand\sdfs{\sigma_{\hat{\pi}}} %
\newcommand\sdrf{\sigma_{\hat{\delta}}} %
\newcommand\covrf{\sigma_{\hat{\delta}\hat{\pi}}} %

\usepackage{graphicx}

\DeclareMathOperator{\corr}{cor}
\DeclareMathOperator{\cov}{cov}
\DeclareMathOperator{\sign}{sign}

\usepackage{float}

\usepackage[
 bibencoding=utf8,%
 maxbibnames=3, %
 minbibnames=1, %
 backend=biber,%
 sortlocale=en_US,%
 style=apa,%
 doi=true,%
 url=true,
 eprint=true%
]{biblatex}

\usepackage{comment}

\usepackage{xcolor}%
\definecolor{webbrown}{rgb}{.6,0,0}%
\usepackage{hyperref} %
\hypersetup{%
  breaklinks = true,%
  colorlinks = true,%
  anchorcolor = webbrown,%
  citecolor = webbrown,%
  filecolor = webbrown,%
  linkcolor = webbrown,%
  menucolor = webbrown,%
  urlcolor= webbrown,%
  citebordercolor= 1 0 0,%
  menubordercolor=1 0 0,%
  urlbordercolor=1 0 0,%
  runbordercolor=1 0 0,%
  pdftitle={One Instrument to Rule Them All},
  pdfauthor={Angrist and Kolesár}
}

\usepackage{cleveref}
\crefname{appsec}{Appendix}{Appendices}
\crefname{appsubsec}{Appendix}{Appendices}
\usepackage[nolist]{acronym}
\begin{acronym}
  \acro{CI}{confidence interval}%
  \acro{OVB}{omitted variable bias}%
  \acro{OLS}{ordinary least squares}%
  \acro{CLT}{central limit theorem}%
  \acro{IV}{instrumental variables}%
  \acro{ATE}{average treatment effect}%
  \acro{LATE}{local average treatment effect}%
  \acro{2SLS}{two-stage least squares}
  \acro{just-ID IV}{just-identified IV with a single endogenous variable}
\end{acronym}

\usepackage{tikz} %

\usepackage{datetime}
\newdateformat{monthyeardate}{\monthname[\THEMONTH] \THEYEAR}
\begin{document}

\title{{One Instrument to Rule Them All:\\The Bias and Coverage of Just-ID IV}\thanks{We thank Ahmet Gulek and Luther Yap for expert research assistance. Thanks also go to Tim Armstrong, Isaiah Andrews, Brigham Frandsen, Guido Imbens, Mike Keane, Dave Lee, Whitney Newey, and Steve Pischke for helpful discussions and insightful comments. Kolesár acknowledges support by the Sloan Research Fellowship and by the National Science Foundation Grant SES-22049356. The views expressed here are our own.}}
\author{Joshua Angrist\and Michal Kolesár}
\monthyeardate%

\maketitle
\begin{abstract}
  We revisit the finite-sample behavior of single-variable just-identified instrumental variables (just-ID IV) estimators, arguing that in most microeconometric applications, the usual inference strategies are likely reliable. Three widely-cited applications are used to explain why this is so. We then consider pretesting strategies of the form $t_{1}>c$, where $t_{1}$ is the first-stage $t$-statistic, and the first-stage sign is given. Although pervasive in empirical practice, pretesting on the first-stage $F$-statistic exacerbates bias and distorts inference. We show, however, that median bias is both minimized and roughly halved by setting $c=0$, that is by screening on the sign of the \textit{estimated} first stage. This bias reduction is a free lunch: conventional confidence interval coverage is unchanged by screening on the estimated first-stage sign. To the extent that IV analysts sign-screen already, these results strengthen the case for a sanguine view of the finite-sample behavior of just-ID IV\@.
\end{abstract}
\thispagestyle{empty}

\setcounter{page}{0}

\clearpage
\section{Introduction}

The heavily over-identified \ac{2SLS} estimates reported in \textcite{angrist90} and \textcite[AK91]{AnKr91} sparked a flood of interest in the finite-sample behavior of \ac{IV} estimators. Kicked off by \textcite{bekker94} and \textcite{bjb95}, attention to bias in \ac{2SLS} estimates with many weak instruments has since become a staple of applied microeconometrics. The fact that the finite-sample distribution of \ac{2SLS} estimates is shifted towards the corresponding \ac{OLS} probability limit is especially worrying. \ac{IV} is often motivated by the belief that \ac{OLS} estimates are compromised by \ac{OVB}. The \ac{IV} analyst hopes, therefore, that when \ac{IV} and \ac{OLS} are close, this signals minimal \ac{OVB} rather than substantial finite-sample bias in the \ac{IV} estimates.

\ac{2SLS} combines multiple instruments in an effort to estimate a single causal effect with acceptable precision. Strikingly, however, \textcite{bjb95} show that in the AK91 specifications interacting quarter of birth dummies with covariates to generate 180 instruments, replacing real quarter of birth dummies with dummies randomly drawn yields \ac{2SLS} estimates and standard errors much like those generated by the real thing.\footnote{This ``fake instruments'' simulation was originally suggested by Alan Krueger. Although not an empirical study, \textcite{bekker94} is likewise motivated by a heavily over-identified specification in \textcite{angrist90} that uses 73 draft lottery dummies plus interaction terms as instruments for Vietnam-era veteran status. This application is featured at the end of Bekker's paper, and, originally, in an Amsterdam bar in 1992, where Paul Bekker first confronted Angrist with claims of finite-sample bias.} But most \ac{IV} studies featuring a single endogenous variable build on a single underlying instrument, such as a dummy for draft-eligibility in \textcite{angrist90} and quarter of birth in AK91. \Ac{just-ID IV} offers a simple, transparent estimation strategy in such cases. Our analysis comes in the wake of renewed interest in the finite-sample properties of \ac{just-ID IV}, an interest reflected in \textcite{AnAr17}, \textcite{lmmp20}, and \textcite{KeNe21}, among others.

We argue here that in typical microeconometric applications of \ac{just-ID IV}, conventional \ac{IV} estimates and $t$-tests are compromised little by failures of the usual asymptotic theory.
Our analysis builds on the (approximate) finite-sample normality of reduced-form and first-stage estimators (in the argot of classical simultaneous equations models, these are both estimated ``reduced forms''). This modeling framework parallels that used in \textcite{ass19} and earlier theoretical investigations of weak instrument problems. The normality of reduced-form estimates is justified by conventional asymptotic reasoning, as well as by the local-to-zero asymptotic sequence used in \textcite{StSt97} and \textcite{StYo05}, in which the first stage shrinks to zero at a rate inversely proportional to the square root of the sample size.

Our setup has two free parameters: the correlation between structural and first-stage residuals (henceforth, ``endogeneity''), and the population first-stage $F$-statistic. This fact lends itself to the construction of easily-interpreted rejection contours characterizing conventional second-stage $t$-tests and confidence interval coverage rates. We see, for example, that for endogeneity less than about $0.76$, 95\% confidence interval coverage is distorted by no more than 5\% for \textit{any} population $F$. This is explained by the fact that, even as median bias increases when the first stage gets weaker, second-stage precision falls (we focus on median bias because the conventional \ac{just-ID IV} estimator has no moments). In contrast with the over-identified case, conventional \ac{just-ID IV} standard errors reflect this, and confidence intervals widen accordingly. This fact keeps interval coverage high unless endogeneity is extraordinarily high.

What range of values for endogeneity is relevant? Three applications are used to calibrate it: AK91, the \textcite[AE98]{AnEv98} \ac{IV} estimates using a dummy for samesex sibships as an instrument for family size, and the \textcite[AL99]{AnLa99} fuzzy regression discontinuity estimates of class size effects on student learning. These studies span a range of \ac{OVB} scenarios, from modest (for most of the AK91 estimates), to substantial (in AE98, where OLS exceeds \ac{IV} by about 50\%), to dramatic (in AL99, where \ac{IV} exceeds small, insignificant OLS estimates by an order of magnitude). Yet, the absolute value of estimated endogeneity is no more than $0.47$ in these applications, and over $0.4$ only for a single specification and sample. Although three examples do not make a theorem, we argue that the features of these studies that limit endogeneity are common to empirical strategies designed to estimate causal effects or to mitigate attenuation bias in models with measurement error.\footnote{The spirit of this argument differs from that in \textcite{StYo05}, which focuses on worst-case rejection rates over all possible endogeneity values. \textcite{lmmp20} and \textcite{KeNe21}, discussed further in \Cref{sec:coverage_contours} below, consider though largely downplay restrictions on endogeneity.}

Evidence on the reliability of conventional \ac{just-ID IV} inference notwithstanding, IV practitioners have come to see weak instruments as problematic in just-identified as well as over-identified models. Responding to \textcite{bjb95,StSt97}, analysts now routinely report first-stage $t$- and $F$-statistics, a practice that's hard to argue with. Yet, \textcite{HaRuWi96} and others since note that requiring, say, a first-stage $F$-statistic greater than 10 when instruments are truly weak often does more harm than good. Requiring first-stage estimates to meet a prespecified cutoff amounts to imposition of a pretest that distorts sampling distributions and makes conventional confidence intervals misleading. This is the IV version of the general pretesting problem highlighted by \textcite{LePo05}.\footnote{\textcite{ass19} survey recent empirical scholarship to demonstrate the empirical relevance of pretesting based on first-sage $F$ in \ac{IV} applications.}

Analyses of pretesting to date apply when the analyst is agnostic as to the sign of the first stage. \textcite{AnAr17} observe, however, that the typical \ac{just-ID IV} scenario includes a theoretical sign restriction. This leads us to consider pretesting strategies that maintain a sign restriction on the population first stage.  Specifically, we examine pretesting rules of the form $t_{1}>c$, where $t_{1}$ is the first-stage $t$-statistic and $c$ is a constant chosen by the analyst. This examination leads to a novel theoretical result: the median bias of \ac{just-ID IV} conditional on $t_{1}>c$ is minimized by choosing $c=0$, that is, by screening on the sign of the estimated first-stage. Moreover, median bias of \ac{just-ID IV} is roughly halved by requiring the estimated first-stage to have the expected (that is, population) sign.

Surprisingly, pre-screening on the estimated first-stage sign is also shown to be virtually costless: rejection contours for a sign-screened estimator differ little from those obtained without screening. The upshot is that sign-screening mitigates the already-modest median bias of \ac{just-ID IV} without degrading coverage. To the extent that such screening is a feature of modern empirical work, reported \ac{IV} estimates reflect the impressively minimal bias characteristic of sign-screened \ac{IV}. Our theoretical results on the bias-minimizing and bias-mitigating consequences of requiring $t_{1}>0$ therefore strengthen the case for a sanguine view of conventional inference for \ac{just-ID IV}.

Finally, the theorem establishing bias mitigation from screening on an estimated first stage sign provides an interesting contrast to \textcite{AnAr17}, which shows how to use a sign restriction on the \textit{population} first stage to construct a mean unbiased \ac{just-ID IV} estimator. We show that conditional on the sign of the \emph{estimated} first stage, this estimator, denoted $\hat{\beta}_{U}$, is no longer unbiased. Rather, $\hat{\beta}_U$ is unbiased by virtue of the fact that it averages two conditional estimators, each biased but in opposite directions. If, as seems likely, empirical sign-screening is endemic in applied microeconomics, this would seem to reduce the appeal of $\hat{\beta}_{U}$ for empirical practice.

The next section details the \ac{just-ID IV} setup assuming normally distributed first-stage and reduced-form estimates and derives an expression for endogeneity in terms of \ac{OLS} \ac{OVB}. 
\Cref{sec:coverage_contours} reviews the relationship between $t$-test rejection rates and the parameters that govern the normal model. 
This section also explains why endogeneity in applied microeconometrics is unlikely to be high enough for conventional IV inference to mislead, and quantifies the length advantage of conventional confidence intervals relative to Anderson-Rubin-based intervals. 
\Cref{sec:signrestrict} presents our theoretical results on first-stage screening. \Cref{sec:conclusion} concludes with a discussion of the implications of our results.
Proofs and details behind numerical calculations appear in the appendix.

\section{Setup}\label{sec:setup}
The sample is assumed to consist of $n$ units indexed by $i$, with data on outcome variable, $Y_{i}$,
a scalar treatment variable, $D_i$, a vector of covariates, $X_i$, and a scalar instrument,
$Z_i$. Population regressions of outcome and treatment on the instrument and
covariates define the reduced form and first stage.  These are written as follows:
\begin{align}
    Y_i & = Z_i\delta + X_i' \psi_1 + u_i,\label{eq:rf}\\
    D_i & = Z_i\pi + X_i'\psi_2 + v_i.\label{eq:fs}
\end{align}
The parameter of interest is $\beta=\frac{\delta}{\pi}$, the ratio of the reduced-form and
first-stage regression coefficients on $Z_{i}$. Provided that the instrument, $Z_i$, satisfies an exclusion restriction and is relevant (i.e.\ $\pi\neq 0$), this parameter captures the causal effect of $D_{i}$ on $Y_{i}$. More generally, if
treatment effects are heterogeneous and a monotonicity condition holds,
$\beta$ is a weighted average of individual causal effects
\parencite{ImAn94,AnIm95}. While treatment effect heterogeneity affects the interpretation of $\beta$, heterogeneity has no bearing on the behavior of the estimators and inference procedures considered in our \ac{just-ID IV} setting.

Let $\hat{\delta}= \sum_{i=1}^{n}\tilde{Z}_{i}Y_{i}/\sum_{i=1}^{n}\tilde{Z}_{i}^{2}$ and $\hat{\pi}=\sum_{i=1}^{n}\tilde{Z}_{i}D_{i}/\sum_{i=1}^{n}\tilde{Z}_{i}^{2}$ denote \ac{OLS} estimates of $\delta$ and $\pi$, where
$\tilde{Z}_{i}$ is the residual from a regression of $Z_{i}$ on $X_{i}$. 
Under mild regularity conditions that allow the errors $(u_i, v_i)$ to be non-normal, heteroskedastic, and serially or cluster-dependent, $(\hat{\delta}, \hat{\pi})$ is consistent and asymptotically normal as $n\to\infty$, with an asymptotic covariance matrix that can be consistently estimated. Importantly, this holds regardless of the strength of the instrument. We therefore follow \textcite{ass19} and earlier analyses of weak instrument problems by assuming this large-sample approximation holds exactly. Specifically, we assume:
\begin{equation}\label{eq:normality}
      \begin{pmatrix}
    \hat{\delta}\\\hat{\pi}
  \end{pmatrix}\sim\mathcal{N}\left(
    \begin{pmatrix}
      \pi\beta\\\pi
    \end{pmatrix}, \Sigma=
    \begin{pmatrix}
        \sdrf^{2} & \covrf\\
        \covrf & \sdfs^{2}
    \end{pmatrix}\right),
\end{equation}
with a known covariance matrix, $\Sigma$. This distributional assumption is implied by the \textcite{StSt97} weak-instrument asymptotic sequence (see
\textcite[Section 3.2]{ass19} for additional discussion and references). Finite-sample results under \cref{eq:normality} can therefore be seen as asymptotic under the \textcite{StSt97} sequence.

\Cref{eq:normality} is our only substantive restriction; this assumption allows us to focus
on the weak instrument problem, separating this from other finite-same problems,
such as the effect of high-leverage observations on the quality of the normal approximation to the distribution of the OLS estimators $(\hat{\delta}, \hat{\pi})$ and the challenge of standard-error estimation with clustered data.\footnote{\textcite{young21iv} discusses these problems in an \ac{IV} context.}
With~\eqref{eq:normality} as foundation, we derive finite-sample properties of the \ac{IV}
estimator:
\begin{equation}\label{eq:iv_estimator}
    \hat{\beta}_{IV} = \frac{\hat{\delta}}{\hat{\pi}},
\end{equation}
and the null rejection rate for the corresponding Wald test. The latter is based on the $t$-statistic centered at the parameter of interest, $\beta$, divided by the estimated \ac{IV} standard error, $\hat{\sigma}_{IV}$:
\begin{equation}\label{eq:wald_CI}
  t_{W}=\frac{\hat{\beta}_{IV}-\beta}{\hat{\sigma}_{IV}}; \qquad
  \hat{\sigma}_{IV}^2=\frac{\sdrf^{2}-2\covrf\hat{\beta}_{IV}+
    \sdfs^{2}\hat{\beta}_{IV}^{2}}{\hat{\pi}^2},
\end{equation}
where $\hat{\sigma}_{IV}^2$ estimates the asymptotic variance of $\hat{\beta}_{IV}$ under standard $n \rightarrow \infty$ asymptotics. The corresponding theoretical variance is ${\sigma}_{IV}^2=(\sdrf^{2}-2\covrf\beta+ \sdfs^{2}\beta^{2})/\pi^2$. In a homoskedastic model with constant causal effects, this simplifies to the familiar formula
\begin{equation*}
  {\sigma}_{IV}^2=\frac{\sigma_{\epsilon}^{2}}{n E[\ddot{Z}_i^2]{\pi}^2},
\end{equation*}
where $\ddot{Z}_i$ is the residual from the population projection of $Z_i$ on $X_{i}$, and $\sigma_{\epsilon}^{2}$ is the variance of the residual in the structural equation,
\begin{equation}\label{eq:struc}
Y_{i}=D_{i}\beta+X_{i}'(\psi_{1}-\psi_{2}\beta)+\epsilon_{i},
\end{equation}
that motivates \ac{IV} estimation in the classic linear set-up (the structural residual is $\epsilon_{i}=u_{i}-v_{i}\beta$).

Given the assumption of a known covariance matrix for the first-stage and reduced-form estimates, both $t_W$ and $\hat{\beta}_{IV}$ depend on the data only through $(\hat{\delta}, \hat{\pi})$. These have distributions determined by the two unknown parameters, $\pi$ and $\beta$. It's illuminating, however, to reparametrize in terms of instrument strength and the degree of endogeneity (a reparametrization adopted in \textcite{StSt97} and \textcite{lmmp20}, among others). The first parameter in this scheme, denoted $E[F]$, is defined as:
\begin{equation*}
E[F]=\pi^{2}/\sdfs^{2}+1.
\end{equation*}
Because $E[F]$ is the expectation of $F=\hat{\pi}^{2}/\sdfs^{2}$, the
$F$-statistic testing $\pi=0$, it's sometimes called the \emph{population
first-stage $F$-statistic}, a term adopted here. Since $\pi$ is a scalar, $E[F]=E[t_{1}]^{2}+1$, where $t_1=\hat{\pi}/\sdfs$ is the first-stage t-statistic.

The second parameter is defined as:
\begin{equation}\label{eq:defrho}
\rho=\corr(\hat{\delta}-\hat{\pi}\beta, \hat{\pi})=
\frac{\sdfs}{
  \sqrt{\sdrf^{2}-2\beta\covrf+\sdfs^{2}\beta^{2}}}\times (\covrf/\sdfs^{2}-\beta).
\end{equation}
With independent heteroskedastic errors, $\rho$ is also given by
$\corr(\ddot{Z}_{i}\epsilon_{i}, \ddot{Z}_{i}v_{i})$.  When, in addition, the errors
$(u_{i}, v_{i})$ are homoskedastic, $\rho=\corr(\epsilon_{i}, v_{i})$, where
$\epsilon_{i}$ is the structural residual in~\eqref{eq:struc}.  We therefore
refer to $\rho$ as (the degree of) \emph{endogeneity}.\footnote{This simplification is obtained using the fact that, under homoskedasticity, the variance of $v_{i}$ is
  $\sigma_{v}^{2}=\sdfs^{2} \cdot n E[\ddot{Z}_i^2]$ and the variance of $\epsilon_i$ is
  $\sigma_{\epsilon}^{2}=(\sdrf^{2}-2\beta\covrf+\sdfs^{2}\beta^{2})\cdot n E[\ddot{Z}_i^2]$, with $\cov(v_i, \epsilon_i)=(\covrf-\beta\sdfs^{2})\cdot n E[\ddot{Z}_i^2]$. The homoskedastic formula for the variance of $\epsilon_i$ also leads yields the simplification of the formula for $\sigma^2_{IV}$ noted above.}

With weak instruments as well as homoskedastic error terms, $\rho$ is proportional to the bias of the \ac{OLS} estimand.
This can be seen by using the first-stage and reduced-form equations to write
the \ac{OLS} slope coefficient, $ \beta_{OLS}$, as follows:
\begin{align}
  \beta_{OLS}&=\frac{E[\ddot{D}_{i} Y_i]}{E[\ddot{D}_i^2]} =\frac{E[\ddot{Z}_i^2]\pi^2\beta+E[u_{i}v_{i}]}{
    E[\ddot{Z}_i^2]\pi^2+\sigma_{v}^2} \nonumber\\
    &=R^2_{p}\beta+
(1-R^2_{p}) \frac{E[u_{i}v_{i}]}{\sigma_{v}^2}; &R^2_p&=\frac{E[\ddot{Z}_i^2]\pi^2}{E[\ddot{Z}_i^2]\pi^2+\sigma_{v}^2}\label{eq:def_partialR2},
\end{align}
where $\ddot{D}_i$ is the residual from a population regression of $D_i$ on $X_i$, and $\sigma^2_{v}=E[v_i^2]$. The weight multiplying $\beta$ in~\eqref{eq:def_partialR2}, denoted $R^2_p$, is the population partial $R^2$ generated by adding the instrument to the first-stage regression. When the instrument is weak, $R^2_p$ is close to zero, and~\eqref{eq:def_partialR2} is approximately
 $E[u_{i}v_{i}]/\sigma_v^2$. The OLS estimand likewise converges to $E[u_{i}v_{i}]/\sigma_v^2$ in the \textcite{StSt97} weak-instrument sequence (which takes $\pi\to 0$). This in turn equals
$\covrf/\sdfs^2$ under homoskedasticity, so the second term on the right-hand side of~\eqref{eq:defrho},
\begin{equation}\label{eq:betaols_definition}
\covrf/\sdfs^{2}-\beta=\beta_{WOLS}-\beta,
\end{equation}
is the weak-instrument OVB of OLS (where we've introduced the notation $\beta_{WOLS}$ for $\covrf/\sdfs^{2})$. Moreover, when $\pi=0$, it follows from~\eqref{eq:normality} that $\beta_{WOLS}-\beta$ is the median bias of $\hat{\beta}_{IV}$, a result requiring no independence or heteroskedasticity assumptions on the errors in~\eqref{eq:rf} and~\eqref{eq:fs}.\footnote{Assumption~\eqref{eq:normality} implies that we can write reduced form and first stage estimates as $\hat{\delta}=\pi\beta+(\covrf/\sdfs)\mathcal{Z}_{\pi}+(\sdrf^2-\covrf^2/\sdfs^2)^{1/2}\mathcal{Z}_{\delta}$ and $\hat{\pi}=\pi+\sdfs \mathcal{Z}_\pi$, where $\mathcal{Z}_{\delta}$ and $\mathcal{Z}_{\pi}$ are independent standard normal variables. When $\pi=0$, therefore, $\hat{\beta}_{IV}=\frac{1}{\sdfs}(\sdrf^2-\covrf^2/\sdfs)^{1/2}(\mathcal{Z}_{\delta}/\mathcal{Z}_\pi)+\beta_{WOLS}$, the median of which is $\beta_{WOLS}$ since $\mathcal{Z}_{\delta}/\mathcal{Z}_\pi$ has a standard Cauchy distribution with zero median.} Thus, $\rho$ also measures endogeneity in the sense that it's proportional to the median bias of the IV estimator when the instrument is irrelevant.

\section{Rejection Rates in Theory and Practice}\label{sec:coverage_contours}

We start by considering $t$-test rejection rates when the null hypothesis is true. For a two-sided $t$-test with level $\alpha$, the rejection rate is the probability that the absolute value of a $t$-statistic, $\abs{t_W}$, exceeds $z_{1-\alpha/2}$, the $1-\alpha/2$ quantile of a standard normal distribution.  The rejection rate of interest can therefore be written:
\begin{equation*}
R_W= P_{E[F], \rho}(\abs{t_W}>z_{1-\alpha/2}),
\end{equation*}
where $P_{E[F], \rho}$ is the distribution of $t_W$ parameterized by $E[F], \rho$. We evaluate $R_W$ by numerical integration, a computation detailed in \Cref{sec:t_test_rejection}.

Summarizing the behavior of a conventional 5\% nominal test, Panel (a) in \Cref{fig:wald} depicts rejection rates for $t_W$ as a contour plot given $\rho$ and $E[F]$. The figure shows that rejection rates substantially exceed the nominal 5\% level only if the instrument is weak (i.e., $E[F]$ is close to $1$) \emph{and} endogeneity is high.
\textcite[Section 3.2]{StYo05} define instruments as weak if the usual 5\% level $t$-test rejects a true null more than 10\% of the time. The figure shows that, as long as $\abs{\rho}<0.76$, rejection rates stay below 10\% regardless of the strength of the first stage.
A stricter standard based on any over-rejection is met as long as $\abs{\rho}<0.565$ \parencite[this cutoff is also noted in][]{lmmp20}.
A simple corollary to these observations, substantiated below by showing that worryingly high values of $\rho$ should be seen as unusual, is that the coverage of conventional nominal 95\% confidence intervals for $\hat{\beta}_{IV}$ is likely to be satisfactory in most applications.

The modest over-rejection seen in \Cref{fig:wald} is explained by a signal feature of \ac{just-ID IV}: the median bias of $\hat{\beta}_{IV}$ rises as the instrument grows weaker, but precision falls apace. The \ac{IV} standard error reflects this lack of precision well enough that, unless endogeneity is egregious, inference is distorted little.
This contrasts with over-identified 2SLS with many weak instruments (as in \textcite{bjb95,bekker94}), where, bias notwithstanding, the usual standard errors for 2SLS remain small enough for the $t$-statistic to be misleading.

Our conclusions here also contrast with those drawn in \textcite{StYo05} and \textcite{lmmp20} regarding the reliability of inference based on a conventional \ac{just-ID IV} $t$-statistic.  Although \textcite{lmmp20} present a similar plot, both studies emphasize worst-case rejection rates over $\rho$, for a given $E[F]$. As can be seen in our \Cref{fig:wald}, this worst-case rejection rate occurs at $\abs{\rho}=1$. In the same spirit, \textcite{KeNe21} highlights simulations showing that conventional \ac{just-ID IV} $t$-tests can be misleading when endogeneity is very high. \Cref{sec:rho_evidence,sec:measurement_error} explain why we are not much concerned with high values of $\rho$.

\textcite{KeNe21} also observe that, since $\hat{\sigma}_{IV}$ and $\hat{\beta}_{IV}$ tend to be negatively correlated when $\rho$ is positive, most false rejections occur when $\hat{\beta}_{IV}>\beta$. This, they argue, militates so strongly against $t_{W}$ that conventional Wald tests are to be avoided even with a first-stage $F$ in the thousands. As we see it, this observation, which is related to asymmetry in the power function of $t_{W}$, does not make conventional frequentist inference unreliable. The conventional standard for reliability of inference is the accuracy of confidence interval coverage, gauged without conditioning on parameter estimates. Our analysis adheres to this standard.\footnote{The \textcite{KeNe21} critique rules out many standard procedures. For instance, consider the conventional $1-\alpha$ confidence interval $[U_{(n)}, U_{(n)}/\alpha^{1/n}]$ for the endpoint of a uniform distribution supported on $[0,\theta]$, where $U_{(n)}$ denotes the largest order statistic in a sample of size $n$. Here all false rejections occur when the minimum variance unbiased estimate $U_{(n)}/(n+1)$ is below $\theta$ (provided $\alpha<1/\exp(1)$).}

\subsection{The Anatomy of Endogeneity}\label{sec:rho_evidence}

We put endogeneity in context using three IV applications.  These are the AK91 study that launched the modern weak instruments literature, the AE98 study using a dummy for samesex sibships (of first- and second-born children) as an instrument for family size, and the AL99 fuzzy regression discontinuity estimates of class size effects. The AE98 and AL99 first-stage $t$-statistics exceed those for AK91 and are arguably out of the zone where an instrument might be considered weak. With a first-stage $t$-statistic of almost 8, the AK91 quarter-of-birth instrument also seems strong enough.  But all three studies can be used to calibrate endogeneity and to document contextual features that constrain it.

\Cref{IVreps} reports key statistics for specifications drawn from each study (some estimates in the table differ slightly from those in the original). The first row in Panel A shows estimates of the economic returns to schooling in the AK91 sample of men born 1920--29. Here, OLS and \ac{IV} estimates equal $0.080$ and $0.072$, respectively. These are close, so endogeneity is small in this case, with an estimated $\rho$ of only $0.043$. Schooling returns estimated in the second AK91 sample, consisting of men born 1930--39, exhibit more OVB\@. In this sample, the \ac{IV} estimate of $0.105$ surprisingly exceeds the OLS estimate of $0.071$ (IV estimation of the returns to schooling is usually motivated by a concern that omitted ability controls causes OLS estimates to be too large). Endogeneity is correspondingly larger at $\rho=-0.175$, but still well outside the danger zone depicted in \Cref{fig:wald}.\footnote{\label{fn:endogeneity_estimates}Endogeneity confidence intervals are computed by inverting the Anderson-Rubin test, and are therefore robust to weak instruments. See \Cref{sec:estimate_rho} for details. In the examples analyzed here, the instruments are not particularly weak, so the bias in estimated endogeneity is negligible and conventional delta-method intervals are similar to those reported in the table.}

The AK91, AE98, and AL99 studies span a range of OVB scenarios, from modest in the first AK91 sample, to substantial in AE98 (where OLS magnitudes consistently exceed \ac{IV} by at least 50\%), to dramatic in AL99 (where \ac{IV} exceeds small, insignificant OLS estimates, mostly by an order of magnitude, and sometimes with a sign flip).  Yet, the magnitude of endogeneity exceeds $0.40$ in only one specification, that for reading scores in the AL99 discontinuity sample (which consists of classes in schools with enrollment near the cutoff that determines class size).  Just-ID IV inference in all three of these studies is therefore unlikely to be compromised by weak instruments.

Although the consistently moderate levels of endogeneity documented in \Cref{IVreps} do not add up to a theorem, these applications have features in common with many IV-driven microeconometric investigations of causal effects.
Specifically, endogeneity in research on causal effects is often capped by the modest size of the causal effects of interest. To make this point, it's helpful to write $\rho$ as a function of OVB\@.  Using \cref{eq:def_partialR2,eq:betaols_definition}, we can express $\rho$ under homoskedasticity as:
\begin{align}\label{eq:defovb}
\rho&=
\frac{\sigma_{v}}{
  \sigma_{\epsilon}}(\beta_{WOLS}-\beta)\\
  &=\frac{\sigma_{v}}{\sigma_{\epsilon}} \left(\frac{\beta_{OLS}-\beta}{1-R^2_{p}} \right)
    \approx \frac{\sigma_{D}}{\sigma_{Y}}(\beta_{OLS}-\beta),\nonumber
\end{align}
where the approximation
$\frac{\sigma_{v}}{\sigma_{\epsilon}(1-R^2_{p})} \approx
\frac{\sigma_{D}}{\sigma_{Y}}$ holds if the explanatory power of the observables
in both the structural and the first-stage equation is low. We can use this expression to compute $\rho$ by replacing $\beta$ with $\hat{\beta}_{IV}$. The relevance of this representation of $\rho$ can be seen in the AE98 estimates of the effects of a third child on weeks worked by women aged 21--35 in the 1980 Census. Here, the difference between OLS and the corresponding \ac{IV} estimate is $-3.42$. Because the first-stage partial R-square ($R^2_{p}$) is close to zero, the term multiplying this, $\frac{\sigma_{v}}{\sigma_{\epsilon}}$, is well-approximated by the ratio of the endogenous variable standard deviation to the dependent variable standard deviation, $\frac{\sigma_{D}}{\sigma_{Y}}$, a ratio of about $0.022$.  The product of these two terms gives $-0.075$, equal to the value of $\rho$ reported in the table for this sample.

\Cref{eq:defovb} suggests a bound on endogeneity motivated by plausible limits to effect size and OVB\@. In the AK91 scenario, for instance, it seems reasonable to assume that the (causal) economic returns to schooling are no more than double the OLS estimand. Under these restrictions, the descriptive statistics in \Cref{IVreps}, which approximate $\frac{\sigma_{v}}{\sigma_{\epsilon}}$ at around 5.2 in this case, suggest $\abs{\rho}$ can be no more than about $0.41$. Although substantial, this is
still below the $0.565$ and $0.76$ values beyond which coverage deteriorates. With $\beta$ bounded below by zero, large magnitudes of $\rho$ require $\beta$ to far exceed $\beta_{WOLS}$. Only when the causal effect of schooling is triple the OLS estimand (so that OLS is \textit{too small} by $0.16$) does the endogeneity danger zone become relevant.\footnote{\textcite{KeNe21} consider bounds on $\rho$ motivated by the view that OLS estimates of schooling returns should exceed causal effects. Although this seems defensible, it's worth noting that the literature surveyed by \textcite{card01lecture} reports many \ac{IV} estimates in excess of the corresponding OLS estimates, a pattern first highlighted by \textcite{lang93}.}

Many microeconometric \ac{IV} applications involve linear probability models in
which causal effects are changes in probabilities. This also has implications for endogeneity. The AE98 estimates of the
effect of the birth of a third child on female labor force participation in
1980, for example, range from roughly $-0.18$ for OLS to $-0.12$ for IV\@. Labor
force participation rates for women with only two children run around 57\%.
Causal effects might therefore be as large as $-0.57$, but no larger, since
probabilities can't be negative. In this case,
$\frac{\sigma_{v}}{\sigma_{\epsilon}}$ is about 1 (again, using standard
deviations in the data rather than residuals), so $\beta_{OLS}-\beta$ can be no
larger than $-0.18+0.57=0.39$, thereby bounding $\rho$ at this value. This generous bound makes no use of
the fact that selection bias is likely to make OLS estimates of family-size effects on female supply too large (in magnitude) rather
than too small. Other applications with Bernoulli outcomes admit similar sorts of bounds.

A related argument, appropriate for models with continuous outcomes, shows endogeneity to be constrained by plausible values for causal effects measured in standard deviation units. This line of reasoning is especially apt for education research where standardized effect sizes are widely reported.  The influential Tennessee STAR class size experiment analyzed in \textcite{krueger99}, for instance, generated a reduction of 7 students per class, roughly one standard deviation of class size in the AL99 data. The STAR experiment yielded treatment effects of about $0.2\sigma$, an impact typical of education interventions deemed to have been effective. At the same time, education researchers often view effect sizes as large as half a standard deviation in the outcome distribution as rare, if not implausible. Using the fact that $\frac{\sigma_{v}}{\sigma_{\epsilon}}$ is about equal to $(1-R^2_{p})$ in the AL99 data, the scenario of a half-standard deviation effect size generated by a one-standard deviation reduction in class size implies $\frac{\sigma_{v}}{\sigma_{\epsilon}}\frac{\beta}{1-R^2_{p}} = -0.5$ on the second line of \cref{eq:defovb}. At the same time, OLS estimates of class size effects in AL99 are mostly zero (as is often found in class size research; see e g., \textcite{hanushek86}), so the magnitude of endogeneity is capped at $0.51$.

Contributing to all three of these empirically-grounded arguments is the fact that endogeneity under homoskedasticity can be split into the difference between two R-squared-like terms:
\begin{equation*}
\rho \approx \frac{\sigma_{D}}{\sigma_{Y}}(\beta_{OLS}-\beta) = \frac{\sigma_{D}}{\sigma_{Y}}\beta_{OLS}-\frac{\sigma_{D}}{\sigma_{Y}}\beta
.
\end{equation*}
The square of the first term, $(\frac{\sigma_{D}}{\sigma_{Y}}\beta_{OLS})^2$, is the variation in the dependent variable accounted for by $D_i$ in an analysis-of-variance for $Y_i$. In microeconometric applications, this sort of $R^2$ term is mostly small, as is the causal analog that determines the square of the second term, $(\frac{\sigma_{D}}{\sigma_{Y}}\beta)^2$. The small size of these two $R^2$ terms limits the magnitude of the difference between them. Consistent with this claim to generality, the many IV estimates collated in \textcite{ChHa08} likewise show modest endogeneity.

It's noteworthy that the bound of $0.41$ derived for the AK91 study depends only on $\beta_{OLS}$, standard deviations $\sigma_{Y}$ and $\sigma_{D}$, and bounds on the causal effect of interest. The resulting calculations therefore seem likely to be relevant for other empirical strategies estimating returns to schooling. As far as details go, our back-of-the-envelope bounds leverage homoskedasticity and the presumption that observables have little explanatory power. In applications where these restrictions are a stretch, \cref{eq:defrho} gives a basis for bounds that apply more generally. Specifically, because $\rho$ is monotone decreasing in $\beta$, plugging bounds on $\beta$ into \cref{eq:defrho} along with estimates of components of the covariance matrix,
$\Sigma$, bounds $\rho$. Estimates of reduced-form and first-stage standard errors are typically readily available, while estimates of $\covrf$ can be
obtained from equation
\cref{eq:covrf_formula} in \Cref{sec:derivations}. For the AK91 study, the more general bound computed using covariance matrix estimates reported in
\Cref{IVreps} leads to a bound on $\rho$ that matches the rough cut. Assuming (as above) that effects of a third child on labor force participation lie between 0 and $-0.57$, and class size effects lie between $0$ and $\frac{\sigma_Y}{2\sigma_D}$, yields upper bounds on $\abs{\rho}$ equal to $0.36$, and $0.57$, respectively. These numbers are likewise close to the corresponding rough-cut bounds of $0.39$ and $0.51$, respectively. 

\subsection{When Measurement Error Motivates IV}\label{sec:measurement_error}

In addition to estimating causal effects, a second major arena for microeconometric \ac{IV} involves models with measurement error. Suppose the regression of interest is
$Y_i=D_i^{*}\beta+X_i'\gamma+\eta_{i}$, where $\eta_i$ is a residual uncorrelated with $(D^*_i, X_i)$ by definition. The regressor $D_i^{*}$ is unobserved; we see only a noisy measure, $D_i=D_i^{*}+e_i$, where the measurement error, $e_i$, is assumed to be classical, that is uncorrelated with $(D_i^{*}, X_i, \eta_i)$. Replacing $D_i^{*}$ with $D_i$ yields the structural equation to be instrumented:
\begin{equation*}
\begin{split}
Y_i &=D_i\beta+X_i'\gamma+(\eta_{i}-e_i\beta) \\
&=D_i\beta+X_i'\gamma+\epsilon_i,
\end{split}
\end{equation*}
where $\epsilon_{i}=\eta_{i}-e_i\beta$ is the structural residual. Given an instrument correlated with $D_i^{*}$ and uncorrelated with $\epsilon_i$, the coefficients of interest are consistently estimated by IV\@. The first stage in this scenario can be written as in~\eqref{eq:fs}, with first-stage residual, $v_i$.

To calibrate endogeneity in this model, note first that, given the classical measurement error assumption, $\cov(v_i, \epsilon_i)=-\sigma^2_e\beta$. Under homoskedasticity, endogeneity squared can therefore be written:
\begin{equation}\label{eq:r}
    \rho^2=\frac{\sigma^4_e\beta^2}{\sigma_v^2\sigma^2_{\epsilon}}
    =\frac{\sigma^4_e\beta^2}{\sigma_v^2(\sigma^2_{\eta}+\beta^2\sigma^2_{e})}
    \leq
    \frac{\sigma^2_e}{\sigma_v^2}
    = \frac{1-r}{1-R^2_{p}},
\end{equation}
where $r=\sigma^2_{\ddot{D}^*}/\sigma^2_{\ddot{D}}$ denotes the reliability (or
signal-to-noise ratio) of mismeasured $D_i$, after partialing out
covariates.\footnote{The first equality in~\eqref{eq:r} follows from the definition of correlation, the middle inequality uses the fact that $\sigma^2_{\eta}$ must be non-negative, and the last equality uses the definition of partial $R^2$ in \cref{eq:def_partialR2}.} Although we can't speak to reliability across all fields, labor economists have collected evidence on the reliability of key variables of interest. 
These include schooling, earnings, hours worked, and hourly wages. Schooling often appears on the right-hand side of wage equations, while earnings, hours, and hourly wages are used in various configurations to estimate labor supply elasticities. 

The \textcite{AnKr99handbook} summary of reliability estimates suggests $r \approx 0.9$ for schooling and $r \approx 0.8$ for earnings, falling to about $0.65-0.75$ for hours worked and hourly wages. The lower end of this range may be more relevant for wage reliability after partialing out covariates or differencing. 
With $r=0.65$ as a reasonably conservative value, we'd need $R_p^2$ equal to at least $0.4$ for $\rho$ to reach $0.76$. 
But $E[F]=\frac{nR^2_p}{1-R^2_p}+1$, so, at this level of first-stage fit, $E[F]$ is nowhere near the trouble zone for any sample size that's empirically relevant.
Of course, reliability can be lower than $0.65$.  Wealth, for instance, is notoriously hard to measure \parencite{SaZu16,szz22}, as is consumption \parencite{bms15}. But neither wealth nor consumption are seen often in the role of a mismeasured endogenous variable to be instrumented. In any case, provided reliability is reasonably high, microeconometric measurement error can be expected to generate parameter combinations for which conventional IV inference is trouble-free.

\subsection{Anderson-Rubin vs. Conventional Confidence Intervals}\label{sec:anderson-rubin-vs}

The \textcite[AR]{AnRu49} statistic for \ac{just-ID IV} offers an alternative to conventional asymptotic inference. AR inference is appealing by virtue of the fact that AR test size is undistorted by weak instruments under the \textcite{StSt97} sequence. Moreover, in the \ac{just-ID IV} context, an AR test is optimal among unbiased tests \parencite{moreira09}. In our setting, the AR statistic can be written:
\begin{equation}\label{eq:t_AR}
t_{AR}=\frac{\hat{\delta}-\hat{\pi}\beta}{\sqrt{\sdrf^{2}-2\covrf\beta+\sdfs^{2}\beta^{2}}}.
\end{equation}
This differs from $t_W$ in that it replaces $\hat{\beta}_{IV}$ with the null value of $\beta$ in the formula for $\hat{\sigma}_{IV}^2$: in the context of the \ac{just-ID IV} model described by \cref{eq:normality}, the square of $t_{AR}$ equals the Lagrange multiplier statistic testing whether $\frac{\delta}{\pi}$ equals $\beta$.\footnote{Since the moment restrictions in the IV model are linear, Proposition 3 in \textcite{NeWe87} implies that $t_{AR}^2$ is also the relevant likelihood ratio statistic. The $t_W$ vs $t_{AR}$ distinction arises solely by virtue of different variance estimators in the denominator, making tests based on these two statistics first-order equivalent.  See \textcite{AnRu49} for the AR statistic in over-identified models with a fixed number of instruments and \textcite{MiSu22} for an adaptation to models with many weak instruments.}  AR tests are also compelling by virtue of the fact that, when testing $\beta=0$, $t_{AR}$ is the $t$-statistic for the associated reduced form. It's hard to imagine a convincing case for statistical significance of a \ac{just-ID IV} estimate when the associated reduced form is statistically indistinguishable from zero. 

The AR test can be inverted to yield a confidence set that guarantees undistorted coverage for any values of $E[F]$ and $\rho$. Why not then default to AR confidence sets? For one thing, AR robustness comes at a cost in precision. AR confidence sets have infinite length when $F\leq z_{1-\alpha/2}^2$ (i.e., less than about 4 for $95\%$ intervals). When $F>z_{1-\alpha/2}^2$, AR intervals are longer than the corresponding conventional intervals. In particular, \Cref{sec:ar_sets} shows that finite AR intervals can be written:
\begin{equation}\label{eq:ar_ci}
    \hat{\beta}_{IV}-r_{AR}
  \pm  \tau_{1}\left(
    \hat{\rho}^{2}(\tau_{1}^{2}-1)
    +1\right)^{1/2}\times z_{1-\alpha/2}\hat{\sigma}_{IV},
\end{equation}
where $\hat{\rho}$ is an endogeneity estimator given in \cref{eq:hatrho}, $\tau_{1}=(1-z_{1-\alpha/2}^{2}/F)^{-1/2}$, and $r_{AR}=\hat{\rho}\hat{\sigma}_{IV}F^{1/2}(\tau_{1}^{2}-1)$.
AR therefore recenters the usual interval at $\hat{\beta}_{IV}-r_{AR}$, while adjusting conventional critical value $z_{1-\alpha/2}$ by $\tau_{1} \left(\hat{\rho}^{2}(\tau_{1}^{2}-1) + 1\right)^{1/2}>1$. 

The AR length penalty is inversely proportional to estimated first-stage strength; this penalty is substantial even for moderately strong instruments. When $F=16$, for instance, the AR interval adjustment factor ranges from 14.7--31.6\%, depending on estimated endogeneity, while the penalty ranges from 8.7--18.2\% when $F=25$. \textcite{lmmp20} develop an appealingly simple alternative robust inference strategy, called $tF$, that adjusts critical values for $t_W$ depending on the value of $F$. Because the $tF$ adjustment is made presuming worst-case endogeneity, however, the penalty here is even larger: $tF$ intervals are 42\% longer than the usual interval when $F=16$, and 25\% longer for $F=25$.\footnote{The $tF$
 interval adjustment is smaller than the AR adjustment for values of $F$ between $z_{1-\alpha/2}^2$ and $6.8$.} 

The bounding arguments illustrated in \Cref{sec:rho_evidence,sec:measurement_error} suggest endogeneity is typically too low for conventional intervals to suffer substantial distortion. In such cases, AR and $tF$ intervals may incur a substantial length penalty while mattering little for coverage. AR and $tF$ intervals are most valuable in applications where endogeneity could plausibly be exceptionally high.

\section{Bias Under a Good Sign}\label{sec:signrestrict}

Having made an empirical case for conventional inference with \ac{just-ID IV}, we add a novel analytical argument showing remarkable bias improvements from screening IV estimates on the sign of the estimated first stage. This argument builds on the idea that \ac{IV} identification strategies are most credible when institutional or theoretical foundations explain the first stage. Such foundations usually imply a sign for $\pi$.  In the AK91 application, for example, the quarter-of-birth first stage arises from the fact that children born later in the year enter school younger, and are therefore constrained by compulsory attendance laws to stay in school longer than those born earlier.  The AE98 samesex instrument for family size is predicated on parents' preference for mixed-sex sibships. The AL99 Maimonides Rule instrument for class size is derived from Israeli regulations that determine class size as a function of enrollment. In these and many other applied micro applications, institutions or preferences sign $\pi$.

\subsection{Sign-Screened Bias and Coverage}\label{sec:sign_screening}

We gauge estimator performance under sign restrictions using median bias since the expectation of a \ac{just-ID IV} estimator is undefined (2SLS moments exists only for over-identified models). Assuming the sign of $\pi$ is known, the theorem below shows that $c=0$ minimizes the median bias of $\hat{\beta}_{IV}$ among screening rules of the form $t_1>c$.

\begin{theorem}\label{theorem:median-bias_c0}
  Consider the model in \cref{eq:normality}, and suppose $\pi>0$. The absolute value of
  the median bias of $\hat{\beta}_{IV}$ conditional on $t_{1}>c$,
  $\abs{\operatorname{median}_{E[F], \rho}(\hat{\beta}_{IV}-\beta\mid
    t_{1}>c)}$, is minimized at $c=0$.
\end{theorem}

\noindent Note that empirical sign-screening yields the greatest bias reduction uniformly over \emph{all} parameter values
$(E[F], \rho)$. In particular, empirical sign-screening reduces median bias
relative to no screening, since the latter sets $c=-\infty$.

For intuition as to why sign-screening is optimal, note first that by virtue of joint normality of first-stage and reduced form estimates in \cref{eq:normality}, the distribution of $\hat{\beta}_{IV}-\beta$ conditional on $t_1$ is normal with a mean and median that can be written:
\begin{equation}\label{eq:conditional_mean}
  E[\hat{\beta}_{IV}-\beta\mid t_{1}]
  =(\beta_{WOLS}-\beta)\frac{t_{1}-E[t_{1}]}{t_{1}}.
\end{equation}
Suppose $\rho>0$, so $\beta_{WOLS}-\beta>0$ is positive. When $t_1$ is positive,
\cref{eq:conditional_mean} implies that conditional median bias is increasing in
$t_1$. Hence, for any $a>0$, screening on $t_1>0$ is better than
screening on $t_1>a$, since IV estimates in samples with $t_1\in [0,a]$ are less biased than estimates
with $t_1>a$. To see why screening on $t_1>0$ is better than screening on $t_1>a$ for $a<0$, note that conditional on negative $t_1$, \cref{eq:conditional_mean} implies that IV bias exceeds that of OLS, because $(t_{1}-E[t_{1}])/t_1>1$. But, as we show in \Cref{theorem:median_bias} below, median bias of IV conditional on $t_1>0$ is smaller than OLS bias.
Inclusion of samples with negative $t_1$ therefore increases median bias. The upshot is an optimal screening cutoff of zero.
\Cref{fig:bias_cond} demonstrates this for selected values of $E[F]$ (numerical computation of median bias is described in \Cref{sec:medi-bias-comparison}). The kinks at zero in the figure reflect the fact that the median of $\hat{\beta}_{IV}$ conditional on $t_{1}$ is discontinuous at zero.

While sign-screening is always salutary, pretesting on $t_{1}>c$ for $c>0$ exacerbates median bias relative to no pretesting unless $E[F]$ is exceedingly small. The figure demonstrates this by marking values of the screening cutoff beyond which screening aggravates bias (these values, determined by bias when $c=-\infty$, are $0.86$ for $E[F]=2$, $0.5$ for $E[F]=3.5$ and $0.38$ for $E[F]=5$). The fact that critical values of $c$ are small in this context explains why pretest rules such as $F>10$ are often counter-productive. Intuitively, large $F$s signal realizations in which the in-sample correlation between instruments and structural errors is largest, exacerbating the median bias of $\hat{\beta}_{IV}$.

The fact that sign-screening affords \textit{substantial} bias reduction is established by a theorem characterizing median IV bias scaled by the weak-IV bias of OLS. Rescaling simplifies bias formulas, while the relationship between conditional and unconditional bias stands without this.\footnote{\textcite{StYo05} use a similar rescaling, focusing on relative mean bias for 2SLS models with over-identifying restrictions.} The theorem below gives a result for worst-case relative bias over $\rho$, which obtains in the limit as $\abs{\rho}\to 0$ (this is not the same as relative bias when $\rho=0$; with no endogeneity, both \ac{IV} and OLS are unbiased, so that relative bias is discontinuous in $\rho$).
The relationship between $\rho$ and relative median bias derived here contrasts with that in \Cref{sec:coverage_contours}, which shows higher endogeneity leads to worse coverage.
This reversal reflects the fact that, although the bias of $\hat{\beta}_{IV}$ increases with endogeneity, OLS OVB increases faster. Worst-case relative bias is characterized by:
\begin{theorem}\label{theorem:median_bias}
  Consider the model in \cref{eq:normality}, and suppose that $\pi>0$. Then, %
  the unconditional relative median bias of $\hat{\beta}_{IV}$ is given by
  \begin{equation}
    \sup_{\rho}\abs*{\frac{\operatorname{median}_{E[F], \rho}(\hat{\beta}_{IV}-\beta)}{\beta_{WOLS}-\beta}}=
    \frac{\phi(E[t_{1}])}{E[t_{1}][\Phi(E[t_{1}])-1/2]+\phi(E[t_{1}])}.\label{eq:bias1}
  \end{equation}
  Moreover, if $E[t_{1}]\geq 0.84$, the relative median
  bias of $\hat{\beta}_{IV}$ conditional on $\hat{\pi}>0$ satisfies
  \begin{equation}
    \sup_{\rho}\abs*{\frac{\operatorname{median}_{E[F], \rho}(\hat{\beta}_{IV}-\beta\mid \hat{\pi}>0)}{\beta_{WOLS}-\beta}}
    =\frac{\phi(E[t_{1}])}{
      E[t_{1}]\Phi(E[t_{1}])+\phi(E[t_{1}])}.\label{eq:bias2}
  \end{equation}
  Equivalently, these expressions give the limit of relative unconditional and conditional median bias as $\abs{\rho}\to 0$.
\end{theorem}

Maintaining the assumptions of the theorem, IV with or without sign-screening has bias below the weak-instrument bias of OLS, since the right-hand sides of both \eqref{eq:bias1} and \eqref{eq:bias2} are in $(0,1)$ as long as $\pi>0$. Note also that the ratio of the two bias expressions in the theorem is close to $0.5$ for all but the smallest values of $E[F]$. Specifically, the ratio of conditional to unconditional median bias is:
\begin{equation*}
  1-\frac{0.5E[t_{1}]}{E[t_{1}]\Phi(E[t_{1}])+\phi(E[t_{1}])}.
\end{equation*}
This quantity is within 1 percentage point of 0.5 once $E[t_{1}]$ greater than about 1.5 since the normal cdf is then close to one and the normal density close to zero.

\Cref{theorem:median_bias} describes worst-case bias over $\rho$. Remarkably, however, the bias reduction from sign-screening varies little with the degree of endogeneity. This is documented in \Cref{fig:bias_ratio}, which plots relative bias as a function of the population first-stage $F$, using shading to mark variation in relative bias as a function of $\rho$ (as in the previous figure, this figure plots numerical calculations detailed in \Cref{sec:medi-bias-comparison}). We see, for example, that for $E[F]$ around $3.5$, the ratio of sign-screened to unconditional bias varies between 0.5--0.52, converging quickly to $0.5$ thereafter.\footnote{\textcite{richardson68} shows that under homoskedasticity, the relative (mean) bias of over-identified \ac{2SLS} relative to (weak-instrument) OLS bias is unrelated to $\rho$.}

The substantial bias reduction generated by sign-screening may seem surprising since wrong-signed first-stage estimates are rare unless $E[F]$ is small. For instance, when $E[F]=3.5$, the probability of a wrong-signed estimate is $P(t_1<0)=\Phi(-E[t_1]) \approx 5.6\%$.  Bias gains from sign-screening arise from 
the fact that, for positive $\rho$, the distribution of $\hat{\beta}_{IV}$
conditional on $t_1<0$ is heavily shifted to the right. Sign-screening therefore discards samples mostly in the far right tail of the IV sampling distribution.  In fact,
\cref{eq:conditional_mean} implies that wrong-signed-conditional median IV bias exceeds OLS bias. So screening yields 
a material improvement in median IV bias even while the events screened out are rare.

While \Cref{theorem:median_bias} establishes the bias-mitigation payoff to sign-screening, \textcite{HaRuWi96} and others show that screening on the first-stage $F$-statistic is a form of pretesting that may degrade inference.
The problem here is that, when $\pi$ is truly zero, large $F$-statistics overstate first-stage strength, leading to
overly optimistic standard errors. And, as noted in the discussion of
\Cref{fig:bias_cond}, conditional bias is aggravated by the fact that large $F$s
signal sample realizations in which the in-sample correlation between instruments and
structural errors is largest, exacerbating the bias of $\hat{\beta}_{IV}$.
Consequently, when instruments are truly weak, pretesting can lead to confidence
intervals with very poor coverage. It's therefore worth investigating whether
empirical sign-screening runs a similar risk. As noted above, wrong-signed first stage estimates are rare when the first stage is nonzero, but pretesting problems are most salient when instruments are indeed weak.

As it turns out, pretesting concerns here are unfounded. By way of evidence on this point, Panel (b) in \Cref{fig:wald} plots rejection contours for a conventional (second-stage) $t$-test conditional on $\hat{\pi}>0$.  That is, the figure plots contours for:
\begin{equation*}
R^{c}_{W}= P_{E[F], \rho}(\abs{t_W}>z_{1-\alpha/2}\mid \hat{\pi}>0).
\end{equation*}
Comparison of the two panels in \Cref{fig:wald} suggests sign-screening affects rejection rates little. For instance, the endogeneity threshold required to keep rejections rates below 10\% is $\abs{\rho}\leq 0.75$, close to the unconditional value of $0.76$ otherwise required for this. This happy finding is explained by the fact that, when the instrument is very weak, sign-screening has two effects. On one hand, screening out wrong-signed first-stage estimates tends to overestimate first-stage strength. At the same time, in contrast to screening on first-stage $F$, the median bias of $\hat{\beta}_{IV}$ is reduced. These two effects are just about offsetting, so that the conditional rejection contours in panel (b) of \Cref{fig:wald} are much like the unconditional contours plotted in panel (a).

\subsection{Theoretical Sign Restrictions Only}

\textcite{AnAr17} introduce an estimator that's unbiased given a sign restriction on the population first-stage coefficient, rather than the sign of estimated $\hat{\pi}$. This unbiased estimator is:
\begin{equation}\label{eq:wted}
  \hat{\beta}_{U}
  \equiv \hat{\tau}(\hat{\delta}-\beta_{WOLS}\hat{\pi})+\beta_{WOLS}
  =t_{1}\mu(t_{1})\hat{\beta}_{IV}+(1-t_{1}\mu(t_{1}))\beta_{WOLS},
\end{equation}
where $\mu(x)=\frac{1-\Phi(x)}{\phi(x)}$ is the Mills' ratio of a standard normal random variable ($\phi$ and $\Phi$ denote the standard normal density and cdf, respectively). Estimator $\hat{\beta}_{U}$ is unbiased because, under normality of first-stage estimates and given $\pi>0$,
\begin{equation*}
    E\left[\frac{\mu(t_{1})}{\sigma_{\hat{\pi}}}\right]=\frac{1}{\pi},
\end{equation*}
that is, $\hat{\tau}\equiv \frac{\mu(t_{1})}{\sigma_{\hat{\pi}}}$
is an unbiased estimator of the reciprocal of $\pi$.  Moreover, $\hat{\delta}-\beta_{WOLS}\hat{\pi}$ and $\hat{\tau}$ are uncorrelated, since
$\beta_{WOLS}$ is the slope in the regression of the estimated reduced form on the estimated first stage. Unbiasedness then follows from the fact that $E[\hat{\delta}-\beta_{WOLS}\hat{\pi}]=(\beta-\beta_{WOLS})\pi$.

$\hat{\beta}_{U}$ has an interesting and counter-intuitive shrinkage interpretation when $t_1>0$.  Observe that
\begin{equation}\label{eq:mills}
0\leq 1-t_{1}\mu(t_{1})\leq \frac{1}{t_{1}^2}
\end{equation}
when $t_1>0$ (this is implied by a Mill's ratio inequality given in \textcite[p.~175]{feller68}). Thus, when the first stage is right-signed, the weights $t_{1}\mu(t_{1})$ in \cref{eq:wted} lie between $0$ and $1$, and $\hat{\beta}_{U}$ shrinks the conventional \ac{IV} estimate towards \ac{OLS}.

The shrinkage interpretation of $\hat{\beta}_{U}$ seems surprising since shrinkage \textit{towards}
OLS increases bias. This fact is reconciled with the unbiasedness of $\hat{\beta}_{U}$ by the following theorem:
\begin{theorem}\label{theorem:bias_betaU}
  Consider the model in~\eqref{eq:normality}, and suppose that $\pi>0$. Then, the mean bias of $\hat{\beta}_{U}$
  conditional on $t_{1}>0$ can be written:
  \begin{equation*}
    E[\hat{\beta}_{U}-\beta\mid t_{1}>0]
    =\frac{0.5e^{-0.5E[t_{1}]^{2}}}{\Phi(E[t_{1}])}(\beta_{WOLS}-\beta)
    ,
  \end{equation*}
  while, conditional on $t_{1}<0$, relative mean bias is:
  \begin{equation*}
  E[\hat{\beta}_{U}-\beta\mid t_{1}< 0]
  =-\frac{0.5e^{-0.5E[t_{1}]^{2}}}{1-\Phi(E[t_{1}])}(\beta_{WOLS}-\beta).
  \end{equation*}
\end{theorem}
\noindent Note that the denominators of these expressions equal the probability $t_1$ is positive and negative, respectively. $\hat{\beta}_{U}$ is therefore unbiased because it averages conditional positive bias when $t_{1}>0$ and conditional negative bias when $t_{1}<0$.

An analyst who is prepared to sign the population first stage seems unlikely to ignore the sign of the estimated first stage. Yet, when it comes to $\hat{\beta}_{U}$, empirical sign-screening results in more bias not less. This would seem to strip $\hat{\beta}_{U}$ of its appeal.
And use of median rather than mean bias to measure performance does not ameliorate this:
\Cref{sec:medi-bias-comparison} shows that the conditional median bias of $\hat{\beta}_{IV}$ is always less than that of $\hat{\beta}_{U}$, and at least 50\% smaller once $E[t_{1}]\geq 1$.\footnote{\textcite{AnAr17} shows numerically that the unconditional median bias of $\hat{\beta}_{U}$ is smaller than that of $\hat{\beta}_{IV}$ when $E[F]$ is small, while this bias ranking reverses for larger $E[F]$. \textcite{AnAr17} notes also that the median absolute deviation of $\hat{\beta}_{U}$ is always smaller than that of $\hat{\beta}_{IV}$. Our numerical calculations indicate that, conditional on the estimated first stage sign, this no longer holds for all parameter values.}

\section{Summary and Conclusions}\label{sec:conclusion}

Assuming reduced-form and first-stage estimates are approximately normally distributed, null rejection rates for conventional $t$-tests in \ac{just-ID IV} models are distorted little unless endogeneity is exceptionally high. A corollary to this fact is good coverage of conventional confidence intervals. Three widely-cited applications, two of which exhibit considerable OLS OVB, are characterized by moderate endogeneity and consequently fall well inside the low-distortion just-ID IV comfort zone. We've argued that these three examples should be seen as representative rather than idiosyncratic: the structure of much applied micro research naturally bounds endogeneity.

We also develop a new theoretical argument alleviating bias concerns in \ac{just-ID IV}. As \textcite{AnAr17} note, the most convincing applications of \ac{just-ID IV} restrict the sign of the first stage.
Unlike \textcite{AnAr17}, however, we impose the same sign restriction on the estimated as well as the theoretical first stage. In contrast to screening on first-stage $F$, which may do more harm than good, empirical sign-screening roughly halves the median bias of the \ac{IV} estimator without degrading coverage. Since most analysts likely impose an estimated first-stage sign screen as a matter of course, the bias reduction sign-conditioning engenders should be reflected in published empirical work.

What practical lesson should we draw from this?  In the context of the AK91, AE98, and AL99 studies, first-stage sign screening adds no action items to the empirical agenda. The first-stage estimates in these applications are robustly right-signed.
In applications with weaker instruments than these, an empirical strategy that begins by examining the first-stage sign would seem to have no downside. Claims for credible causal evidence requires more than this, however. In AK91, for instance, the quarter-of-birth story holds water because schooling can be seen to move sharply up and down with quarter of birth as predicted by compulsory attendance laws, across 30 birth cohorts in three data sets, and because graduate degree completion, which should be changed little by compulsory attendance, moves little with quarter of birth. This sort of coherence contributes as much or more than statistical significance to first stage strength.

\printbibliography
\newpage

\begin{figure}[htp!]
  \centering\input{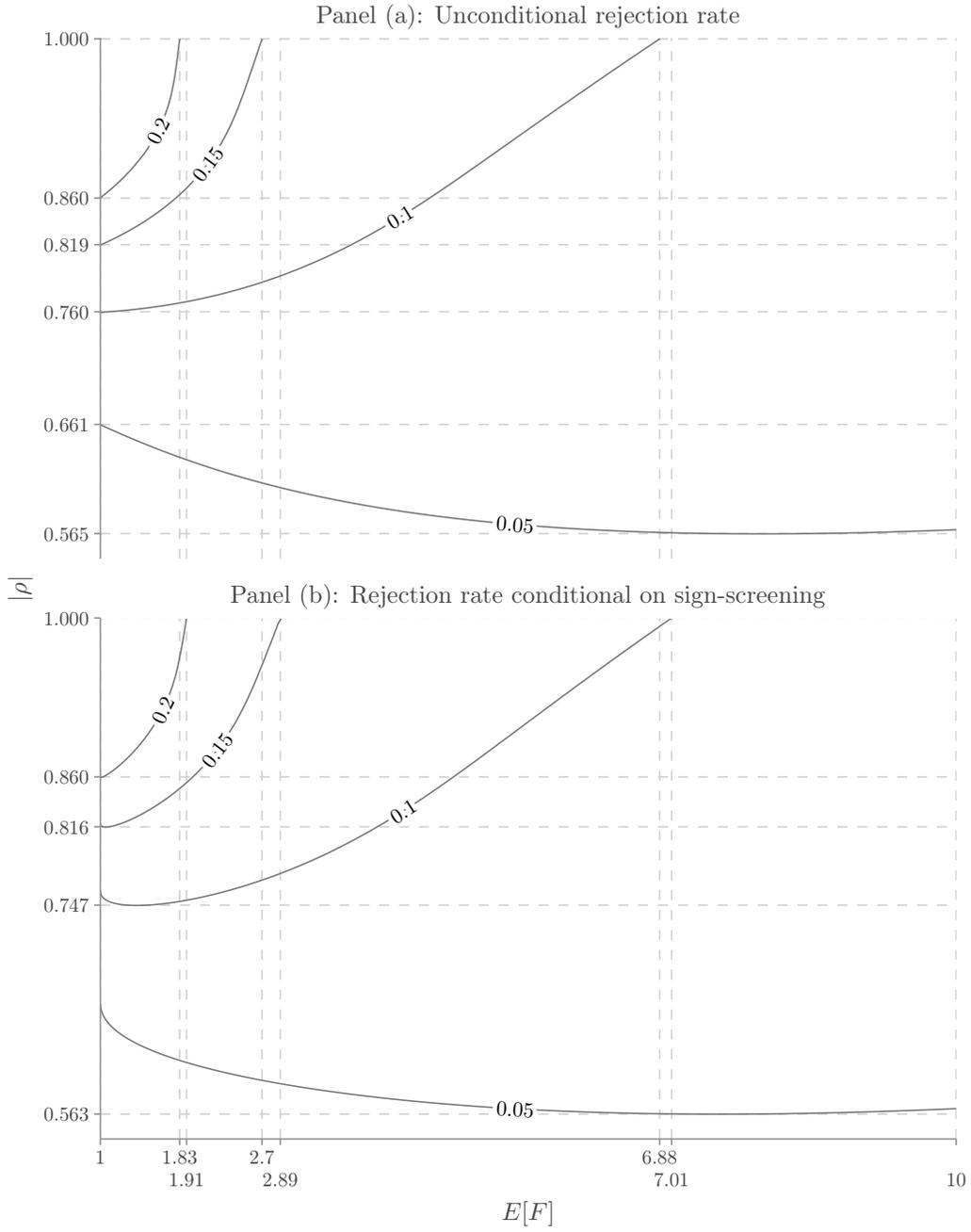}
  \caption{Contour plot of the rejection rate of conventional $t$-test with nominal level $\alpha=0.05$ as function of $E[F]$ and $\rho$. Panel (a) plots the unconditional rejection rate $R_{W}$. Panel (b) plots the rejection rate $R^{c}_{W}$ conditional on $\hat{\pi}>0$. See \Cref{sec:t_test_rejection} for computational details.}\label{fig:wald}
\end{figure}

\begin{figure}[tp]
  \centering\input{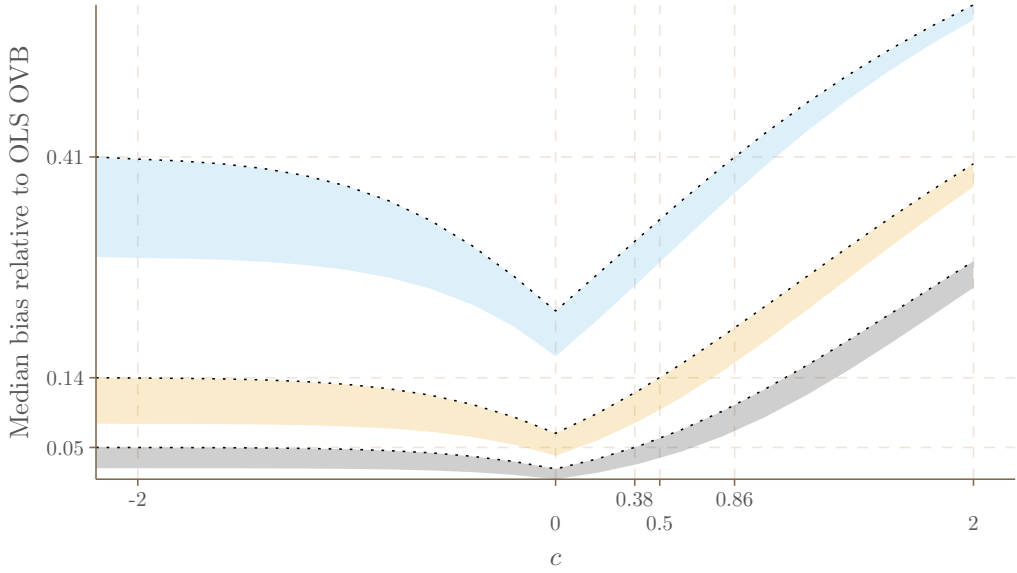}
  \caption{Relative median bias of $\hat{\beta}_{IV}$ conditional on $t_{1}>c$. The figure plots
    $\operatorname{median}(\hat{\beta}_{IV}-\beta\mid
    t_{1}>c)/\abs{\beta_{WOLS}-\beta}$ as a function of test cutoff $c$ for select
    values of $E[F]$. Shaded areas cover the range of variation in
    relative bias over possible values of $\rho$. Blue: $E[F]=2$, orange:
    $E[F]=3.5$, gray: $E[F]=5$. Dotted lines denote limiting relative
    bias as $\rho\to 0$. For each value of $E[F]$, horizontal gridlines mark the value of this relative bias when $c=-\infty$, and vertical gridlines mark the cutoff value for which screening on $t_1>c$ increases bias relative to no screening. See \Cref{sec:medi-bias-comparison} for computational details.}\label{fig:bias_cond}
\end{figure}

\begin{figure}[tp]
  \centering\input{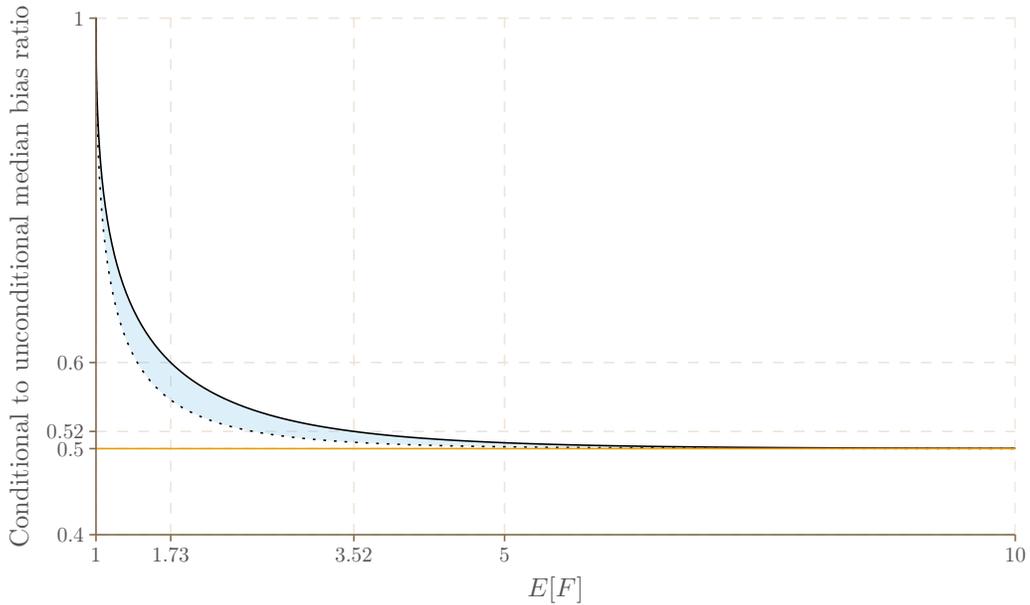}
  \caption{Median bias of $\hat{\beta}_{IV}$ conditional on $t_{1}>0$ relative to unconditional median bias.
  The solid line plots the bias ratio when $\rho=1$; the dotted line denotes the limit of the bias ratio as $\rho\to 0$.
  The blue shaded area covers the range of the bias ratio over possible values of $\rho$. The solid orange line marks a reference line at 0.5. See \Cref{sec:medi-bias-comparison} for computational details.}\label{fig:bias_ratio}
\end{figure}

\clearpage

\begin{landscape}
  \footnotesize\renewcommand*{\arraystretch}{1.02}%
  \newcolumntype{Y}{>{\centering\arraybackslash}X}
  \begin{threeparttable}
    \caption{Estimates and Endogeneity in Three \ac{IV} Applications.}\label{IVreps}
    \begin{tabular}{@{}l cc@{}c @{}S@{}S@{}S@{}S c@{}}
    \toprule
    &&&&\multicolumn{5}{c}{Estimates}\\
    \cmidrule(rl){5-9}
    & Outcome & Treatment & Instrument & \multicolumn{1}{c}{OLS} & \multicolumn{1}{c}{$\hat{\pi}$} & \multicolumn{1}{c}{$\hat{\delta}$} & \multicolumn{1}{c}{IV} & \multicolumn{1}{c}{$\hat{\rho}$}\\
    Sample & (1) & (2) & (3) & \multicolumn{1}{c}{(4)} & \multicolumn{1}{c}{(5)} & \multicolumn{1}{c}{(6)}& \multicolumn{1}{c}{(7)} & \multicolumn{1}{c}{(8)}\\
      \midrule
      \csname @@input\endcsname table1.tex
      \bottomrule
    \end{tabular}
    \begin{tablenotes}
    \item\footnotesize\emph{Notes}: This table reports \ac{IV} and OLS estimates replicating the AK91, AE98, and AL99 studies discussed in the text.
    For each study, the table reports IV and OLS estimates from multiple samples, as well as the corresponding first-stage estimates, $\hat{\pi}$, reduced-form estimates, $\hat{\delta}$, and estimates of the endogeneity parameter, $\hat{\rho}$. Standard errors appear in parentheses. These are robust for AK91 and AE98, and clustered on school for AL99. Confidence intervals for $\rho$, reported below parameter estimates, are computed as described in \Cref{sec:estimate_rho}.
    Standard deviations for the outcome, treatment, and instrument are reported in columns 1--3 in brackets.
    \end{tablenotes}
  \end{threeparttable}
\end{landscape}

\begin{appendices}
\crefalias{section}{appsec}
\crefalias{subsection}{appsubsec}
\numberwithin{equation}{section}
\numberwithin{figure}{section}
\section{Derivations and Proofs}\label{sec:derivations}

The appendix uses the notation
$\tilde{\beta}_{IV}=(\hat{\beta}_{IV}-\beta)/\abs{\beta_{WOLS}-\beta}$ and
$\tilde{\beta}_{U}=(\hat{\beta}_{IV}-\beta)/\abs{\beta_{WOLS}-\beta}$ to denote
the IV and the unbiased estimator, after centering and scaling by the weak-IV OVB of OLS\@.
Also, we let $\omega=\rho/\sqrt{1-\rho^{2}}$.

\subsection{Estimating \texorpdfstring{$\rho$}{rho}}\label{sec:estimate_rho}

We estimate $\rho$ (defined in \cref{eq:defrho}) using first-stage and \ac{IV} estimates, and the associated first-stage, reduced-form and \ac{IV} standard errors. To see how this works, rewrite \cref{eq:wald_CI} as:
\begin{equation}\label{eq:covrf_formula}
  \covrf=
  \frac{\sdfs^{2}\hat{\beta}_{IV}^{2} - \hat{\pi}^{2}\hat{\sigma}^{2}_{IV}+\sdrf^{2}}{2\hat{\beta}_{IV}}.
\end{equation}
With this in hand for $\covrf$, endogeneity can be computed as the sample analog of \cref{eq:defrho}, replacing $\beta$ with $\hat{\beta}_{IV}$.  The resulting endogeneity estimator is:
\begin{equation}\label{eq:hatrho}
    \hat{\rho}=\frac{\sdfs}{\abs{\hat{\pi}} \hat{\sigma}_{IV}}\times
    (\covrf/\sdfs^{2}-\hat{\beta}_{IV}).
\end{equation}
Under the normal model in \cref{eq:normality}, this estimator depends on the data only through $\hat{\beta}_{IV}$, with the derivative given by $\partial\hat{\rho}/\partial{\hat{\beta}_{IV}}=\hat{\sigma}_{IV}^{-1}\cdot (\hat{\rho}^2-1)/\abs{t_{1}}$. Hence, the delta-method standard error for $\hat{\rho}$ is simply $(1-\hat{\rho}^2)/\abs{t_{1}}$. 

Paralleling concerns with finite-sample coverage of the usual confidence interval for $\beta$, we might worry that confidence intervals for $\rho$ based on delta-method standard errors suffer from undercoverage if endogeneity is high and the instruments are weak. We therefore compute confidence sets for $\rho$ by inverting the AR statistic. Specifically, denote the AR confidence set by $[\beta_\ell, \beta_u]$ when this is finite. Since $\rho$ is monotone decreasing in $\beta$, this leads to a confidence set for $\rho$ that can be written $[\varrho(\beta_u), \varrho(\beta_\ell)]$, where $\varrho(\beta)=\frac{\sdfs}{
  \sqrt{\sdrf^{2}-2\beta\covrf+\sdfs^{2}\beta^{2}}}\times
(\covrf/\sdfs^{2}-\beta)$. When the AR confidence set takes the form $(-\infty,
\beta_{\ell}]\cup [\beta_u, \infty)$, the confidence set for $\rho$ is $[-1,\varrho(\beta_u)]\cup
[\varrho(\beta_{\ell}), 1]$.

\subsection{AR confidence sets for \texorpdfstring{$\beta$}{beta}}\label{sec:ar_sets}

The AR confidence set consists of all points $\beta_{0}$ that are not rejected
by the AR test. These points must therefore satisfy the inequality
$ z_{1-\alpha/2}^{2}\geq\frac{(\hat{\delta}-\hat{\pi}\beta_{0})^{2}}{
  \sdrf^{2}-2\covrf\beta_{0}+\sdfs^{2}\beta_{0}^{2}}$. Letting
${\Delta}_{0}=\beta_{0}-\hat{\beta}_{IV}$, and using \cref{eq:wald_CI,eq:hatrho},
we can write this inequality as
\begin{equation*}
  (F-z_{1-\alpha/2}^{2})\Delta_{0}^{2}
  +
  2z_{1-\alpha/2}^{2}\hat{\rho}F^{1/2}\hat{\sigma}_{IV}\Delta_{0}
  -z_{1-\alpha/2}^{2} F\hat{\sigma}_{IV}^{2} \leq 0.
\end{equation*}
Solving this quadratic inequality, we obtain that when $F>z_{1-\alpha/2}^{2}$,
the confidence interval for $\beta-\hat{\beta}_{IV}$ has endpoints given by
\begin{equation*}
- (\tau_{1}^{2}-1)\hat{\rho}F^{1/2}\hat{\sigma}_{IV}\pm
\tau_{1}
z_{1-\alpha/2}\hat{\sigma}_{IV}\sqrt{(\tau_{1}^{2}-1)\hat{\rho}^{2}+
1
}.
\end{equation*}
It follows that the confidence interval for $\beta$ is given by \cref{eq:ar_ci}.

\subsection{\texorpdfstring{$t$}{t}-Test Rejection Rates}\label{sec:t_test_rejection}

This section writes the rejection probabilities of the $t$-test as an integral
indexed by $(E[F], \rho)$. \textcite{StYo05} use Monte Carlo methods to compute
unconditional rejection probabilities in a similar setup. The calculation
described here is much faster. More importantly, it allows us to
easily compute both unconditional rejection rates and rejection rates
conditional on sign-screening.

Using \cref{eq:defrho}, and the fact that $\beta_{WOLS}-\beta$ and $\rho$ have
the same sign, we may write $t_{AR}$ as
\begin{equation}\label{eq:ar_test}
  t_{AR}%
  =\frac{(\hat{\delta}-\hat{\pi}\beta)\abs{\rho}}{\sdfs\abs{\beta_{WOLS}-\beta}}.
\end{equation}
Consequently,
\begin{equation}\label{eq:rewrite-IV}
  \tilde{\beta}_{IV}=\frac{\hat{\delta}-\beta\hat{\pi}}{\sdfs t_{1}\abs{\beta_{WOLS}-\beta}}
  =\frac{t_{AR}}{\abs{\rho}t_{1}}.
\end{equation}
Thus,
\begin{equation}\label{eq:rewrite-wald}
  t_{W}%
  =\frac{\sign(t_{1})t_{AR}}{
    \sqrt{\frac{\sdrf^{2}/\sdfs^{2}-2\beta_{WOLS}\beta+\beta^{2}}{(\beta_{WOLS}-\beta)^{2}}
      \rho^{2} +\frac{t_{AR}^{2}}{t_{1}^{2}}-2\rho\frac{t_{AR}}{t_{1}}
    } }=\frac{\sign(t_{1})t_{AR}}{
    \sqrt{1 +t_{AR}^{2}/t_{1}^{2}-2\rho t_{AR}/t_{1}
    } }
\end{equation}
where the first equality uses \cref{eq:rewrite-IV} and the definition of
$\beta_{WOLS}$, and the second equality uses \cref{eq:defrho}. This expression
for $t_{W}$ implies that conditional on $t_{1}$, the rejection region
$\{\abs{t_{W}}\geq z_{1-\alpha/2}\}$ is quadratic in $t_{AR}$.
Solving this quadratic inequality implies that the rejection region is given by
\begin{equation*}
 t_{AR}\in \begin{cases}
\emptyset & \text{if $t_{1}^{2}\leq (1-\rho^{2})z^{2}_{1-\alpha/2}$,} \\
[a_{1}, a_{2}] & \text{if $ (1-\rho^{2})z^{2}_{1-\alpha/2}\leq t_{1}^{2}\leq z^{2}_{1-\alpha/2}$.}\\
(-\infty, a_{2})\cup (a_{1}, \infty) & \text{if $t_{1}^{2}\geq z^{2}_{1-\alpha/2}$}
\end{cases}
\end{equation*}
where
\begin{align*}
  a_{1}&=\frac{\rho z^{2}_{1-\alpha/2}t_{1}-
         \abs{t_{1}}z_{1-\alpha/2}\sqrt{t_{1}^{2}-(1-\rho^{2})z^{2}_{1-\alpha/2}
         }
         }{z^{2}_{1-\alpha/2}-t_{1}^{2}},\\
  a_{2}&=\frac{\rho z^{2}_{1-\alpha/2}t_{1}+
         \abs{t_{1}}z_{1-\alpha/2}\sqrt{t_{1}^{2}-(1-\rho^{2})z^{2}_{1-\alpha/2}
         }
         }{z^{2}_{1-\alpha/2}-t_{1}^{2}}.
\end{align*}
Note that $\corr(t_{AR}, t_{1})=\rho$, so that
\begin{equation}\label{eq:ar_given_z}
  P(t_{AR}\leq x\mid t_{1})=\Phi((x-\rho (t_{1}-E[t_{1}]))/\sqrt{1-\rho^{2}}).
\end{equation}
Thus, conditional on $t_{1}$, the rejection probability is
given by
\begin{equation}\label{eq:conditional_rejection}
  \begin{split}
    P(\abs{t_{W}}\geq z_{1-\alpha/2}\mid t_{1})&= (P(t_{AR}\leq a_{2}\mid
    t_{1})-P(t_{AR}\leq a_{1}\mid t_{1}))\1{t_{1}^{2}\geq
      z_{1-\alpha/2}^{2}(1-\rho^{2})}\\
    &\qquad +\1{t_{1}^{2}\geq z_{1-\alpha/2}^{2}}\\
    & = f(t_{1};E[t_{1}], \rho)\1{t_{1}^{2}\geq(1-\rho^{2})
      z_{1-\alpha/2}^{2}}+\1{t_{1}^{2}\geq z_{1-\alpha/2}^{2}},
  \end{split}
\end{equation}
where
\begin{equation*}
  f(t_{1};E[t_{1}], \rho)=
  \Phi\left(\frac{a_{2}-\rho
      (t_{1}-E[t_{1}])}{\sqrt{1-\rho^{2}}}\right)- \Phi\left(\frac{a_{1}-\rho
      (t_{1}-E[t_{1}])}{\sqrt{1-\rho^{2}}}\right).
\end{equation*}
Since $t_{1}\sim\mathcal{N}(E[t_{1}], 1)$, the rejection probability conditional
on $t_{1}\geq c$ is therefore given by
\begin{equation*}
  P(\abs{t_{W}}\geq z_{1-\alpha/2}\mid t_{1}\geq c)=\frac{
    \int_{c}^{\infty} (\1{\frac{t_{1}^{2}}{1-\rho^{2}}\geq z_{1-\alpha/2}}f(t_{1};E[t_{1}], \rho)+\1{t_{1}^{2}\geq z_{1-\alpha/2}^{2}})
    \phi(t_{1}-E[t_{1}])dt_{1}}{
    \Phi(E[t_{1}]-c)}.
\end{equation*}
The unconditional rejection probability $R_{W}$ obtains by setting $c=-\infty$.
The rejection probability conditional on sign screening,
$R^{c}_{W}$, obtains by setting $c=0$. The rejection contours in
\Cref{fig:wald} evaluate the above expression
as a function of $(\rho, E[t_{1}])$ by numerical integration.

\subsection{Proof of Theorem~\ref{theorem:median-bias_c0}}

The distribution of $\tilde{\beta}_{IV}$ conditional on $t_{1}$ can be written
as
\begin{equation}\label{eq:tilde_beta_distro}
    P_{\omega, E[t_{1}]}(\tilde{\beta}_{IV}\leq x\mid t_{1})
     =\begin{cases} P(t_{AR} \leq \abs{\rho}t_{1}x\mid t_{1}) =
         \Phi(\omega[E[t_{1}]-(1-\sign(\omega) x) t_{1}])
         & \text{if $t_{1}\geq 0$,} \\
         P(t_{AR}\geq \abs{\rho}t_{1} x\mid t_{1})=
         \Phi(-\omega[E[t_{1}]-(1-\sign(\omega) x) t_{1}])& \text{if $t_{1}<0$,}
    \end{cases}
\end{equation}
where the first equality uses~\cref{eq:rewrite-IV}, and the second equality
follows from \cref{eq:ar_given_z}. Observe that since
$P_{-\omega, E[t_{1}]}(\tilde{\beta}_{IV}\leq x\mid t_{1})=1-P_{\omega, E[t_{1}]}(\tilde{\beta}_{IV} \leq
-x\mid t_{1})$, the distribution is symmetric in $\omega$. It therefore
suffices to consider $\omega>0$.

Let $p_{c}(x)=P_{\omega, E[t_{1}]}(\tilde{\beta}_{IV}\leq x\mid t_{1}>c)$ denote the distribution of $\tilde{\beta}_{IV}$ conditional on $t_{1}>c$ (this shorthand notation ignores the dependence on $\omega$ and $E[t_{1}]$).
By~\cref{eq:tilde_beta_distro},
\begin{equation*}
  p_{c}(x)=
\begin{cases}
  \frac{1}{\Phi(E[t_{1}]-c)}
  \int_{c-E[t_{1}]}^{\infty}
  \Phi\left(y(z)\right)
  \phi(z)dz & \text{if $c\geq 0$,} \\
  \frac{1}{\Phi(E[t_{1}]-c)}\left[\int_{c-E[t_{1}]}^{-E[t_{1}]}\Phi(
  -y(z))\phi(z)dz+
  \int_{-E[t_{1}]}^{\infty}\Phi(y(z))\phi(z)dz\right] & \text{if $c<0$,}
\end{cases}
\end{equation*}
where $y(z)=\omega[xE[t_{1}]-(1-x)z]$. Observe that for all $c$, the conditional
median of $\tilde{\beta}_{IV}$, denoted $m_{c}=m_{c}(\omega)$, is smaller than
$1$. In particular, for $c\geq 0$,
\begin{equation}\label{eq:median_lt1}
  p_{c}(1)
  = \frac{\Phi\left(\omega E[t_{1}]\right)}{\Phi(E[t_{1}]-c)}
  \int_{c-E[t_{1}]}^{\infty}
    \phi(z)dz=\Phi(\omega E[t_{1}])\geq 1/2,
\end{equation}
while if $c<0$,
\begin{multline*}
  p_{c}(1)
  =
  \frac{1}{\Phi(E[t_{1}]-c)}\left[
    \Phi(E[t_{1}]-c)-\Phi(E[t_{1}])
    +\Phi(\omega E[t_{1}])(-\Phi(E[t_{1}]-c)+2\Phi(E[t_{1}]))
  \right]\\
  \geq\frac{1}{\Phi(E[t_{1}]-c)} \left[\Phi(E[t_{1}]-c)-\Phi(E[t_{1}])
    +\frac{1}{2}(2\Phi(E[t_{1}])-\Phi(E[t_{1}]-c))\right]
  =\frac{1}{2}.
\end{multline*}
We now show that the $m_{c}$ is minimized at $c=0$. If $c\geq 0$, by Leibniz rule,
\begin{equation*}
  \frac{\partial p_{c}(x)}{\partial c}
  =
  \frac{\phi(E[t_{1}]-c)}{\Phi(E[t_{1}]-c)}\left[p_{c}(x)
  -  \Phi\left(\omega[E[t_{1}]-(1-x)c]\right)
  \right],
\end{equation*}
which is negative for $x\leq 1$, since it follows
from~\cref{eq:tilde_beta_distro} that
$p_{c}(x)\leq \frac{1}{\Phi(E[t_{1}]-c)} \int_{c-E[t_{1}]}^{\infty}
\Phi(\omega[E[t_{1}]-(1-x)c])
\phi(z)dz=\Phi\left(\omega[E[t_{1}]-(1-x)c]\right)$. Therefore, $m_{c}$ is
decreasing for positive $c$.

Now consider $c<0$. From \cref{eq:tilde_beta_distro}, it follows that
\begin{equation*}
  p_{c}(x)=
  \frac{1}{\Phi(E[t_{1}]-c)}\left[\int_{c-E[t_{1}]}^{-E[t_{1}]}\Phi(
    \omega[-E[t_{1}]+(1-x)(z+E[t_{1}])])\phi(z)dz+\Phi(E[t_{1}])p_{0}(x)\right].
\end{equation*}
Suppose $p_{0}(m_{c})< 1/2$. Then it follows from the preceding display that for
$x\leq 1$,
\begin{multline*}
  \frac{1}{2}< \frac{1}{\Phi(E[t_{1}]-c)}\left[\int_{c-E[t_{1}]}^{-E[t_{1}]}\Phi(
    \omega[-E[t_{1}]+(1-x)(z+E[t_{1}])])\phi(z)dz+\Phi(E[t_{1}])\frac{1}{2}\right]\\
  \leq \frac{1}{\Phi(E[t_{1}]-c)}\left[\frac{1}{2}
    \int_{c-E[t_{1}]}^{-E[t_{1}]}\phi(z)dz+\Phi(E[t_{1}])\frac{1}{2}\right]=\frac{1}{2},
\end{multline*}
where the second inequality uses the fact that
$\Phi(\omega[-E[t_{1}]+(1-x)(z+E[t_{1}])])\leq \Phi(-E[t_{1}]\omega)\leq \Phi(0)$
over the range of integration. Hence, $p_{0}(m_{c})\geq 1/2$, which implies that
$m_{0}\leq m_{c}$. By the proof of \Cref{theorem:median_bias}
$m_{0}\geq 0$. Thus, setting $c=0$ minimizes $\abs{m_{c}}$ for all $c$, as
claimed.

\subsection{Proof of Theorem~\ref{theorem:median_bias}}\label{sec:proof-median_bias}

The proof begins by characterizing the distribution of $\tilde{\beta}_{IV}$
conditional on $t_{1}>0$. The previous proof establishes that the conditional median for this distribution is less than $1$, so it
suffices to consider $p_{0}(x)$ for $x\leq 1$. By the mean value theorem, for
some $\tilde{\omega}=\tilde{\omega}(x, \omega)\in[0,\omega]$,
\begin{multline*}
p_{0}(x)= \Phi\left(0\right)
  +\frac{\omega}{\Phi(E[t_{1}])} \int_{-E[t_{1}]}^{\infty} ((x-1)z+xE[t_{1}])
  \phi\left(\tilde{\omega} ((x-1)z+xE[t_{1}])\right)\phi(z)dz\\
  =\frac{1}{2} +\frac{\omega}{\tilde{\omega}^{2}(1-x)\Phi(E[t_{1}])} \int_{-\infty}^{\tilde{\omega}E[t_{1}]} y
  \phi\left(y\right)\phi(a+by)dy,
\end{multline*}
where the second line uses the change of variables
$y=\tilde{\omega}xE[t_{1}]-\tilde{\omega}(1-x)z$, and we let $a=xE[t_{1}]/(1-x)$,
$b=-\frac{1}{\tilde{\omega}(1-x)}$. By line 111 of Table 1 in \textcite{owen80},
\begin{equation}\label{eq:owen111}
  \int
  x\phi(x)\phi(a+bx)=\frac{\phi(a/t)}{t^{2}}\left[
-\phi(tx+ab/t)-\frac{ab}{t}\Phi(tx+ab/t)\right], \quad t=\sqrt{1+b^{2}}.
\end{equation}
Applying this result to the preceding display then yields
\begin{multline*}
p_{0}(x) =\frac{1}{2}
  +\frac{\omega}{\tilde{\omega}^{2}(1-x)\Phi(E[t_{1}])} \frac{\phi(a/t)}{1+b^{2}}\left[
    -\phi\left(t\tilde{\omega}E[t_{1}]+ab/t\right)
    -\frac{ab}{\sqrt{1+b^{2}}}\Phi(t\tilde{\omega}E[t_{1}]+ab/t)\right]\\
  = \frac{1}{2} +\frac{\omega}{\Phi(E[t_{1}])}
  \frac{\phi(a/t)(1-x)}{\tilde{\omega}^{2}(1-x)^{2}+1}\left[
    \frac{x}{\abs{1-x}}\frac{E[t_{1}]}{\tilde{g}(x, \tilde{\omega})}
    \Phi(E[t_{1}]
    g(x, \tilde{\omega})) -\phi\left(E[t_{1}] g(x, \tilde{\omega})\right) \right] ,
\end{multline*}
where
$g(x, \tilde{\omega})=
  \frac{\tilde{\omega}^{2}\abs{1-x}+\sign(1-x)}{
  \sqrt{\tilde{\omega}^{2}(1-x)^{2}+1}}$, and
$\tilde{g}(x, \tilde{\omega})=\sqrt{\tilde{\omega}^{2}(1-x)^{2}+1}$. When evaluated at
the conditional median, $m_{0}$, the expression in square brackets must equal zero by definition of
the median. Therefore, $m_{0}>0$, and since we also know
from~\cref{eq:median_lt1} that $m_{0}<1$, the conditional median must satisfy
\begin{equation}\label{eq:conditional_median_bias}
  m_{0}=\frac{1}{\frac{E[t_{1}]}{\tilde{g}(m_{0}, \tilde{\omega}(m_{0}, \omega))}
    \frac{\Phi\left(E[t_{1}] g(m_{0}, \tilde{\omega}(m_{0}, \omega))\right)}{
      \phi(E[t_{1}]
      g(m_{0}, \tilde{\omega}(m_{0}, \omega)))}+1},
\end{equation}
We have
\begin{equation*}
  \frac{E[t_{1}]}{\tilde{g}}\frac{\Phi\left(E[t_{1}] g\right)}{
    \phi(E[t_{1}]
    g)}\geq  \frac{E[t_{1}]}{\tilde{g}}\frac{\Phi\left(E[t_{1}] \tilde{g}\right)}{
    \phi(E[t_{1}]\tilde{g})}, \quad\text{and}\quad
  \frac{E[t_{1}]}{\tilde{g}}\frac{\Phi\left(E[t_{1}] \tilde{g}\right)}{
    \phi(E[t_{1}]\tilde{g})}\geq
  E[t_{1}]
  \frac{\Phi\left(E[t_{1}]\right)}{
    \phi(E[t_{1}])} \quad\text{if $E[t_{1}]\geq 0.84$}.
\end{equation*}
Here the first inequality follows because $\Phi(x)/\phi(x)$ is increasing in
$x$, and $g\geq \tilde{g}$, and the second inequality follows because
$\frac{\Phi(x)}{x\phi(x)}$ is increasing for $x\geq 0.84$, and $\tilde{g}\geq 1$.
Therefore,
\begin{equation*}
  m_{0}\leq \frac{\phi(E[t_{1}])}{E[t_{1}] \Phi(E[t_{1}])+\phi(E[t_{1}])}=\lim_{\omega\downarrow 0}m_{0}(\omega),
\end{equation*}
where the equality follows since the right-hand side of
\cref{eq:conditional_median_bias} converges to $\frac{\phi(E[t_{1}])}{E[t_{1}]
  \Phi(E[t_{1}])+\phi(E[t_{1}])}$ as $\omega\to 0$.

We now prove the claims concerning the unconditional distribution of
$\tilde{\beta}_{IV}$. By arguments in the proof of
\Cref{theorem:median-bias_c0}, the median is smaller than $1$, so it suffices to
consider $p_{-\infty}(x)$ for $x\leq 1$. By arguments as in the conditional case,
\begin{equation*}
  \begin{split}
    p_{-\infty}(x)&= \frac{1}{2} +\frac{\omega}{\tilde{\omega}^{2}(1-x)}
    \left[\int_{-\infty}^{\tilde{\omega}E[t_{1}]}
      y\phi\left(y\right)\phi(a+by)dy
      -\int_{\tilde{\omega}E[t_{1}]}^{\infty}y\phi\left(y\right)\phi(a+by)dy\right]\\
    &= \frac{1}{2}
    +\frac{\omega}{\tilde{\omega}^{2}(1-x)}\frac{\phi(a/t)}{1+b^{2}}\left[
      -2\phi(t\tilde{\omega}E[t_{1}]+ab/t)-2\frac{ab}{t}\Phi(t\tilde{\omega}E[t_{1}]+ab/t)
      +\frac{ab}{t}\right]\\
    &= \frac{1}{2}
    +\frac{\omega}{\tilde{\omega}^{2}(1-x)}\frac{\phi(a/t)}{1+b^{2}}\left[
      -2\phi(E[t_{1}] g(\tilde{\omega}, x ))
      +2\frac{x}{1-x}\frac{E[t_{1}]}{\tilde{g}} \Phi(E[t_{1}] g(\tilde{\omega},
      x)) -\frac{x}{1-x}\frac{E[t_{1}]}{\tilde{g}}\right].
  \end{split}
\end{equation*}
Here the first line follows by the mean value theorem, where
$\tilde{\omega}=\tilde{\omega}(x, \omega)\in[0, \omega]$, the second line uses~\cref{eq:owen111}, and
the last line follows by algebra. When evaluated at $x=m_{-\infty}$, the expression in
square brackets must equal zero by definition of the median. Therefore,
$m_{-\infty}>0$, and it must satisfy
\begin{equation}\label{eq:unconditional_median_bias}
  m_{-\infty}= \frac{1} {
    \frac{E[t_{1}]}{\tilde{g}}\frac{\Phi(E[t_{1}]
      g)-1/2}{\phi(E[t_{1}] g)}+1
  }
\end{equation}
Now,
\begin{equation*}
  \frac{E[t_{1}]}{\tilde{g}}\frac{\Phi(E[t_{1}]
      g)-1/2}{\phi(E[t_{1}] g)}\geq
  \frac{E[t_{1}]}{\tilde{g}}\frac{\Phi(E[t_{1}]
      \tilde{g})-1/2}{\phi(E[t_{1}] \tilde{g})}
\geq E[t_{1}]\frac{\Phi(E[t_{1}])-1/2}{\phi(E[t_{1}])}.
\end{equation*}
Here the first inequality follows because $\Phi(x)/\phi(x)$ is increasing in
$x$, and $g\geq \tilde{g}$, and the second inequality follows because
$\frac{\Phi(x)-1/2}{x\phi(x)}$ is increasing for $x>0$. As a result,
\begin{equation*}
  m_{-\infty}\leq \frac{\phi(E[t_{1}])}{E[t_{1}](\Phi(E[t_{1}])-1/2+\phi(E[t_{1}]))}=\lim_{\omega\downarrow 0}m_{-\infty}(\omega),
\end{equation*}
where the equality follows since the right-hand side of
\cref{eq:unconditional_median_bias} converges to
$\frac{\phi(E[t_{1}])}{E[t_{1}](\Phi(E[t_{1}])-1/2+\phi(E[t_{1}]))}$ as $\omega\to 0$.

\subsection{Proof of Theorem~\ref{theorem:bias_betaU}}\label{sec:proof-betaU}
We may write
\begin{equation}\label{eq:tildebetaU}
  \tilde{\beta}_{U}=t_{1}\mu(t_{1})\tilde{\beta}_{IV}+(1-t_{1}\mu(t_{1}))\sign(\rho)
  =\mu(t_{1})\frac{t_{AR}}{\abs{\rho}}+(1-t_{1}\mu(t_{1}))\sign(\rho)
\end{equation}
where the first equality follows from \cref{eq:wted}, and the fact that $\beta_{WOLS}-\beta$ and $\rho$ have the
same sign, and the second equality applies \cref{eq:rewrite-IV}.

Since $E[t_{AR}\mid t_{1}]=\rho(t_{1}-E[t_{1}])$, the relative bias conditional
on $t_{1}$ is given by
\begin{equation*}
  E[\tilde{\beta}_{U}\mid t_{1}]=\sign(\rho)\left[
    1-E[t_{1}]\mu(t_{1})\right].
\end{equation*}
By arguments analogous to those in the proof of Lemma 2.1 in \textcite{AnAr17},
we have
\begin{equation*}
  \begin{split}
    E[t_{1}]E[\mu(t_{1})\mid t_{1}>0] &
    =\frac{E[t_{1}]}{\Phi(E[t_{1}])}\int_{t=0}^{\infty}\frac{1-\Phi(t)}{\phi(t)}\phi(t-E[t_{1}])dt
    =
    \frac{e^{-\frac{E[t_{1}]^{2}}{2}}}{\Phi(E[t_{1}])}\int_{t=0}^{\infty}(1-\Phi(t)) E[t_{1}] e^{E[t_{1}] t}dt\\
    &=\frac{e^{-E[t_{1}]^{2}/2}}{\Phi(E[t_{1}])}\left\{\left[e^{E[t_{1}]
          t}(1-\Phi(E[t_{1}]))\right]_{t=0}^{\infty}
      +\int_{t=0}^{\infty}\phi(t)e^{E[t_{1}] t}dt \right\}\\
    &= \frac{1}{\Phi(E[t_{1}])}\left[-\frac{1}{2}e^{-E[t_{1}]^{2}/2}
      +\int_{t=0}^{\infty}\phi(t-E[t_{1}]) dt\right]
    =-\frac{1}{2}\frac{e^{-E[t_{1}]^{2}/2}}{\Phi(E[t_{1}])}+1,
  \end{split}
\end{equation*}
where the first line uses the definition of the Mills' ratio, the second line uses
integration by parts, and the third follows by completing the square. It
therefore follows that
\begin{equation*}
  \frac{E[\hat{\beta}_{U}-\beta\mid t_{1}>0]}{\beta_{WOLS}-\beta}=
  \frac{1}{2}\frac{e^{-E[t_{1}]^{2}/2}}{\Phi(E[t_{1}])}
  .
\end{equation*}
The second claim follows by an analogous argument.

\subsection{Median Bias Comparisons}\label{sec:medi-bias-comparison}

To evaluate the relative median bias of $\hat{\beta}_{IV}$ as a function of both $E[F]$
and $\rho$ conditional on $t_{1}\geq c$, we first evaluate its distribution
\begin{equation}\label{eq:median_iv_formula}
  P(\tilde{\beta}_{IV}\leq x\mid t_{1}\geq c;\rho, E[t_{1}])
  =\frac{1}{\Phi(E[t_{1}]-c)}
  \int_{c-E[t_{1}]}^{\infty}f_{IV}(z;x, E[t_{1}], \omega)\phi(z)dz
\end{equation}
by numerical integration. Here we use the formula
$f_{IV}(z;x, \rho, E[t_{1}])=\Phi(\omega[\sign(z+E[t_{1}])E[t_{1}]-(1-x)\abs{z+E[t_{1}]}])$
from \cref{eq:tilde_beta_distro} for the cdf conditional on $z=t_{1}-E[t_{1}]$.
We then numerically solve for the median. The
shaded regions in \Cref{fig:bias_cond} correspond to the range of the absolute value
of the relative median bias as $\rho$ varies between $-1$ and $1$. Similarly, the shaded regions in \Cref{fig:bias_ratio} show how the median bias conditional on $t_1\geq 0$ relative to the unconditional median bias (that sets $c=-\infty$) varies with $\rho$.

To compare the relative median bias to that of $\hat{\beta}_{U}$, it suffices to
consider $\rho>0$, since the distributions of $\hat{\beta}_{U}$ and
$\hat{\beta}_{IV}$ are symmetric in $\rho$. By \cref{eq:tildebetaU}, it follows
that for $t_{1}>0$,
\begin{equation*}
  P(\tilde{\beta}_{U}\leq x\mid t_{1}; \omega)=
  P\left(\tilde{\beta}_{IV}
    \leq x-(1-x)\frac{(1-t_{1}\mu(t_{1}))}{t_{1}\mu(t_{1})}\;\Big|\; t_{1}; \omega\right),
\end{equation*}
which for $x<1$ is smaller than $P(\tilde{\beta}_{IV}\leq x\mid t_{1}; \omega)$. Since
the median of $\tilde{\beta}_{IV}$ conditional on $t_{1}>0$ is smaller than $1$,
it follows that the conditional median bias of $\tilde{\beta}_{IV}$ is always
smaller than that of $\tilde{\beta}_{U}$.

To compare the relative magnitudes of the median biases, we compute the relative
median bias of $\tilde{\beta}_{U}$ analogously to that of $\tilde{\beta}_{IV}$,
except we replace $f_{IV}$ in \cref{eq:median_iv_formula} with
$f_{U}(z;x, E[t_{1}],
\rho)=\Phi(\omega[E[t_{1}]-(1-\sign(\omega)x)/\mu(E[t_{1}]+z)])$ (it follows from
\cref{eq:tildebetaU,eq:ar_given_z} that this is the cdf $\tilde{\beta}_{U}$
conditional on $z=t_{1}-E[t_{1}]$). We then compute the ratio
$\operatorname{median}_{E[t_{1}], \rho}(\tilde{\beta}_{U}\mid
t_{1}>0)/\operatorname{median}_{E[t_{1}], \rho}(\tilde{\beta}_{IV}\mid t_{1}>0)$
of the median biases on a fine grid of values of $(\rho, E[t_{1}])$. This ratio
is greater than $2$ if $E[F]\geq 2$, and greater than $3$ if $E[F]\geq 3$,
regardless of the value of $\rho$. Likewise, comparison of the ratio of the
conditional and unconditional median IV bias,
$\operatorname{median}_{E[t_{1}], \rho}(\tilde{\beta}_{IV}\mid
t_{1}>0)/\operatorname{median}_{E[t_{1}], \rho}(\tilde{\beta}_{IV})=
\operatorname{median}_{E[t_{1}], \rho}(\hat{\beta}_{IV}-\beta\mid
t_{1}>0)/\operatorname{median}_{E[t_{1}], \rho}(\hat{\beta}_{IV}-\beta)$ shows
that the ratio lies between $0.5$ and $0.525$ for $E[t_{1}]\geq 1.5$, regardless
of the value of $\rho$.

\end{appendices}
\end{document}